%% file: main.tex
\documentclass[10pt,twocolumn,letterpaper]{article}

\usepackage{cvpr}
\usepackage{times}
\usepackage{epsfig}
\usepackage{graphicx}
\usepackage{amsmath}
\usepackage{amssymb}

\usepackage{mystyle}
\usepackage{CJKutf8}

\usepackage[pagebackref=true,breaklinks=true,colorlinks,bookmarks=false]{hyperref}

\cvprfinalcopy 

\def\cvprPaperID{2724} 

\ifcvprfinal\fi
\begin{document}

\title{Deep Parametric Shape Predictions using Distance Fields}

\author{Dmitriy Smirnov$^1$ \quad Matthew Fisher$^2$ \quad Vladimir G. Kim$^2$ \quad Richard Zhang$^2$ \quad Justin Solomon$^1$\\
$^1$Massachusetts Institute of Technology \quad $^2$Adobe Research
}

\maketitle

\begin{abstract}
    Many tasks in graphics and vision demand machinery for converting shapes into consistent representations with sparse sets of parameters; these representations facilitate rendering, editing, and storage. When the source data is noisy or ambiguous, however, artists and engineers often manually construct such representations, a tedious and potentially time-consuming process. While advances in deep learning have been successfully applied to noisy geometric data, the task of generating parametric shapes has so far been difficult for these methods. Hence, we propose a new framework for predicting parametric shape primitives using deep learning.  We use distance fields to transition between shape parameters like control points and input data on a pixel grid. We demonstrate efficacy on 2D and 3D tasks, including font vectorization and surface abstraction.
\end{abstract}

\input{sections/introduction.tex}
\input{sections/related_work.tex}
\input{sections/preliminaries.tex}
\input{sections/method.tex}
\input{sections/2d.tex}
\input{sections/3d.tex}
\input{sections/conclusion.tex}

\section{Acknowledgement}

The authors acknowledge the generous support of Army Research Office grant W911NF1710068, Air Force Office of Scientific Research award FA9550-19-1-031, of National Science Foundation grant IIS-1838071, National Science Foundation Graduate Research Fellowship under Grant No. 1122374, from an Amazon Research Award, from the MIT-IBM Watson AI Laboratory, from the Toyota-CSAIL Joint Research Center, from a gift from Adobe Systems, and from the Skoltech-MIT Next Generation Program.

{\small
\bibliographystyle{ieee_fullname}
\bibliography{bibliography}
}

\clearpage

\input{sections/supplementary.tex}
\end{document}

%% file: sections/introduction.tex
\vspace{-0.152in}
\section{Introduction}

The creation, modification, and rendering of parametric shapes, such as in vector graphics, is a fundamental problem of interest to engineers, artists, animators, and designers. Such representations offer distinct advantages. By expressing shapes as collections of primitives, we can easily apply transformations and render at arbitrary resolution while storing only a sparse representation. Moreover, generating parametric representations that are \emph{consistent} across inputs enables us to learn common underlying structure and estimate correspondences between shapes, facilitating tools for retrieval, exploration, style/structure transfer, and so on.

It is often useful to generate parametric models from data that do not directly correspond to the target geometry and contain imperfections or missing parts. This can be an artifact of noise, corruption, or human-generated input; often, an artist intends to create a precise geometric object but produces one that is ``sketchy'' and ambiguous. Hence, we turn to machine learning methods, which have shown success in inferring structure from noisy data.

Convolutional neural networks (CNNs) achieve state-of-the-art results in vision tasks such as image classification~\cite{krizhevsky2012imagenet}, segmentation~\cite{long2015fully}, and image-to-image translation~\cite{pix2pix2017}. CNNs, however, operate on \emph{raster} representations. Grid structure is fundamentally built into convolution as a mechanism for information to travel between network layers. This structure is leveraged to optimize GPU performance. Recent deep learning pipelines that output vector shape primitives have been significantly less successful than pipelines for analogous tasks on raster images or voxelized volumes.

A challenge in applying deep learning to parametric geometry is the combination of Eulerian and Lagrangian representations. CNNs process data in an \emph{Eulerian} fashion, applying fixed operations to a dense grid; Eulerian shape representations like indicator functions come as values on a fixed grid.  Parametric shapes, on the other hand, use sparse sets of parameters like control points to express geometry. In contrast to stationary Eulerian grids, this \emph{Lagrangian} representation moves with the shape.  Mediating between Eulerian and Lagrangian geometry is key to any learning pipeline for the problems above, a task we consider in detail.

\begin{figure}[t!]
    \centering
    \includegraphics[width=\linewidth]{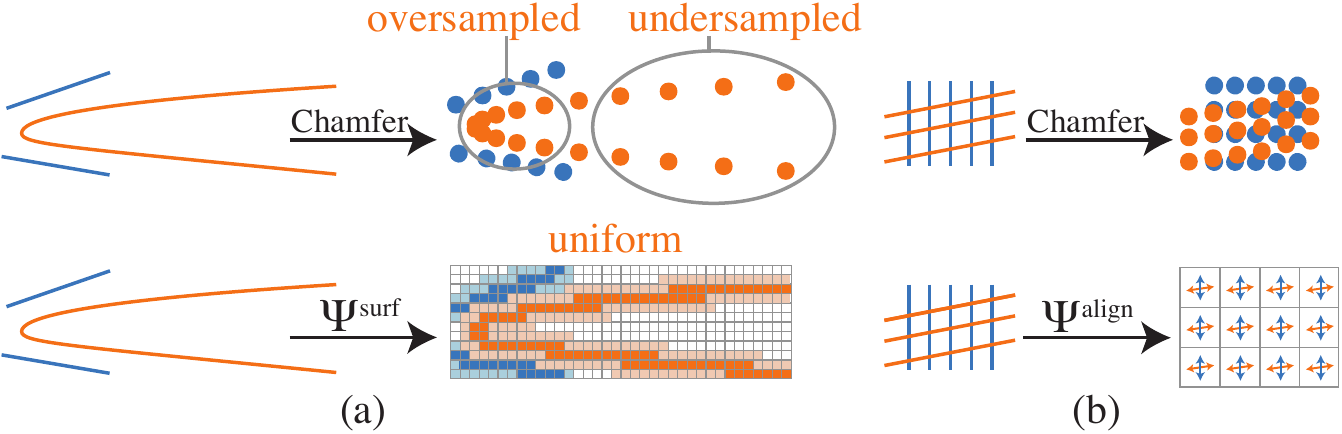}
    \caption{Drawbacks of Chamfer distance (above) fixed by our losses (below). In (a), sampling uniformly in the parameter space of a B\`ezier curve (orange) yields oversampling at the high-curvature area, resulting in a low Chamfer distance to the segments (blue). Our method yields a spatially uniform representation. In (b), two sets of nearly-orthogonal line segments have near-zero Chamfer distance despite misaligned normals. We explicitly measure normal alignment.\vspace{-0.2in}}
    \label{fig:chamfer-drawbacks}
\end{figure}

We propose a learning framework for predicting parametric shapes, addressing the aforementioned issues. By analytically computing a distance field to the primitives during training, we formulate an Eulerian version of Chamfer distance, a common metric for geometric similarity \cite{tulsiani2017learning, fan2017point, liu2004hand, groueix2018atlasnet}.  Our metric does not require samples from the predicted or target shapes, eliminating artifacts that emerge due to nonuniform sampling. Additionally, our distance field enables alternative loss functions that are sensitive to specific geometric qualities like alignment. We illustrate the advantages of our method over Chamfer distance in Figure~\ref{fig:chamfer-drawbacks}.

We apply our new framework in 2D to a diverse dataset of fonts, training a network that takes in a raster image of a glyph and outputs a collection of B\'ezier curves. This effectively maps glyphs onto a common set of parameters that can be traversed intuitively. We use this embedding for font exploration and retrieval, correspondence, and interpolation in a completely self-supervised setting, without need for human labelling or annotation.

We also show that our approach works in 3D. With surface primitives in place of curves, we perform abstraction on ShapeNet \cite{shapenet2015}, outputting parametric primitives to approximate each input. Our method can produce consistent shape segmentations, outperforming state-of-the-art deep cuboid fitting of Tulsiani et al.\ \cite{tulsiani2017learning} on semantic segmentation.

\vspace{-0.15in}
\paragraph*{Contributions.} We present a technique for predicting parametric shapes from 2D and 3D raster data, including:
\begin{itemize}
    \item a \emph{general distance field loss function} motivating several self-supervised losses based on a common formulation;
    \item application to 2D \emph{font glyph vectorization}, with application to correspondence, exploration, retrieval, and repair;
    \item application to 3D \emph{surface abstraction}, with results for different primitives and constructive solid geometry (CSG) as well as application to segmentation.
\end{itemize}

%% file: sections/related_work.tex
\section{Related Work}
\paragraph*{Deep shape reconstruction.} Reconstructing geometry from one or more viewpoints is crucial in applications like robotics and autonomous driving  \cite{fuentes2015visual,seitz2006comparison,su2015multi}. Recent deep networks can produce point clouds or voxel occupancy grids given a single image \cite{fan2017point,choy20163d}, but their output suffers from fixed resolution.

Learning signed distance fields defined on a voxel grid \cite{dai2017shape,stutz2018learning} or directly \cite{Park_2019_CVPR} allows high-resolution rendering but requires surface extraction; this representation is neither sparse nor modular. Liao et al.\ address the rendering issue by incorporating marching cubes into a differentiable pipeline, but the lack of sparsity remains problematic, and predicted shapes are still on a voxel grid \cite{liao2018deep}.

Parametric shapes offer a sparse, non-voxelized solution. Methods for converting point clouds to geometric primitives achieve high-quality results but require supervision, either relying on existing labeled data \cite{niu2018im2struct,mo2019structurenet,gao2019deepspline} or prescribed templates \cite{ganapathi2018parsing}. Groueix et al.\ output primitives at any resolution, but their primitives are not naturally parameterized or sparsely represented \cite{groueix2018atlasnet}. Genova et. al.\ propose to represent geometry as isosurfaces of axis-aligned Gaussians \cite{genova2019shape}. Others \cite{groueix2018atlasnet,sun2019abstraction,Paschalidou2019CVPR} develop tailored primitives but use standard Chamfer distance as the loss objective. We demonstrate and address the issues inherent in Chamfer distance.

\vspace{-0.1in}
\paragraph*{Font exploration and manipulation.}
Designing or even finding a font can be tedious using generic vector graphics tools. Certain geometric features distinguish letters from one another across fonts, while others distinguish fonts from one another. Due to these difficulties and the presence of large font datasets, font exploration, design, and retrieval have emerged as challenging problems in graphics and learning.

Previous exploration methods categorize and organize fonts via crowdsourced attributes \cite{o2014exploratory} or embed fonts on a manifold using purely geometric features \cite{campbell2014learning,Balashova18}. Instead, we leverage deep vectorization to automatically generate a sparse representation for each glyph. This enables exploration on the basis of general shape rather than fine detail.

Automatic font generation methods usually fall into two categories. Rule-based methods \cite{suveeranont2010example,phan2015flexyfont} use engineered decomposition and reassembly of glyphs into parts. Deep learning approaches \cite{azadi2018multi,upchurch2016z} produce raster images, with limited resolution and potential for image-based artifacts, making them unfit for use as glyphs. We apply our method to edit existing fonts while retaining vector structure and demonstrate vectorization of glyphs from noisy partial data.

\vspace{-0.1in}
\paragraph*{Parametric shape collections.} As the number of publicly-available 3D models grows, methods for organizing, classifying, and exploring models become crucial. Many approaches decompose models into modular parametric components, commonly relying on prespecified templates or labeled collections of specific parts \cite{kim2013learning,shen2012structure,ovsjanikov2011exploration}. Such shape collections prove useful in domain-specific applications in design and manufacturing \cite{schulz2017retrieval,umetani2012guided}.  Our deep learning pipeline allows generation of parametric shapes to perform these tasks. It works quickly on new inputs at test time and is generic, handling a variety of modalities without supervision and producing different output types.

%% file: sections/preliminaries.tex
\section{Preliminaries}
\label{sec-prelim}

Let $A, B \subset \R^n$ be two measurable shapes. Let $X$ and $Y$ be two point sets sampled uniformly from $A$ and $B$. The \emph{directed Chamfer distance} between $X$ and $Y$ is
\begin{equation}\label{eq:chamferdir}
    \Ch_\mathrm{dir}(X,Y) = \frac1{|X|} \sum_{x \in X} \min_{y \in Y} \|x-y\|_2^2,
\end{equation}
and the \emph{symmetric Chamfer distance} is defined as 
\begin{equation}\label{eq:chamfer}
    \Ch(X,Y) = \Ch_\mathrm{dir}(X,Y) + \Ch_\mathrm{dir}(Y,X).
\end{equation}
\noindent These were proposed for computational applications in \cite{borgefors1984distance} and have been used as a loss function assessing similarity of a learned shape to ground truth in learning \cite{tulsiani2017learning, fan2017point, liu2004hand, groueix2018atlasnet}.

To relate our proposed loss to Chamfer distance, we define \emph{variational directed Chamfer distance} as
\begin{equation}
    \Ch_\mathrm{dir}^\mathrm{var}(A,B) = \frac1{\operatorname{Vol(A)}} \int_A \inf_{y \in B} \|x-y\|^2_2\,\dint V(x),
\end{equation}
with \emph{variational symmetric Chamfer distance} $\Ch(A,B)^\mathrm{var}$ defined analogously, extending \eqref{eq:chamferdir} and \eqref{eq:chamfer} to smooth objects.

If points are sampled uniformly, under relatively weak assumptions, $\Ch(X,Y)\!\to\!0$ iff $A\!=
\!B$ as the number of samples grows, making it a reasonable shape matching metric.
Chamfer distance, however, has fundamental drawbacks:
\begin{itemize}[leftmargin=*]
    \item It is highly dependent on the sampled points and sensitive to non-uniform sampling, as in Figure~\ref{fig:chamfer-drawbacks}a.
    \item It is agnostic to normal alignment. As in Figure~\ref{fig:chamfer-drawbacks}b, Chamfer distance between a dense set of vertical lines and a dense set of horizontal lines approaches zero.    
    \item It is slow to compute. For each $x$ sampled from $A$, it is necessary to find the closest $y$ sampled from $B$, a quadratic-time operation when implemented na\"ively. Efficient structures like $k$-d trees are not well-suited to GPUs.
\end{itemize}
Our method does not suffer from these disadvantages.

%% file: sections/method.tex
\section{Method}
\label{sec-method}

We introduce a framework for formulating loss functions suitable for learning parametric shapes in 2D and 3D; our formulation not only generalizes Chamfer distance but also leads to stronger loss functions that improve performance on a variety of tasks. We start by defining a general loss on distance fields and propose two specific losses.

\subsection{General Distance Field Loss}

Given $A,B\subseteq\R^n$, let $\dd_A, \dd_B : \R^n \to \R_+$ measure distance from each point in $\R^n$ to $A$ and $B$, respectively, $\dd_A(x)\eqdef \inf_{y\in A} \|x-y\|_2$. In our experiments, $n \in \{2,3\}.$
Let $S \subseteq \R^n$ be a bounded set with $A,B \subseteq S$. We define a \emph{general distance field loss} as
\begin{equation}
\label{gdf-loss}
    \LL_\Psi[A, B] = \frac{1}{\operatorname{Vol}(S)} \int_{x\in S} \Psi_{A,B}(x)\,\dint V(x),
\end{equation}
for some measure of discrepancy $\Psi$. Note that we represent $A$ and $B$ only by their respective distance functions, and the loss is computed over $S$.

Let $\Phi \in \R^p$ be a collection of parameters defining a shape $S_\Phi \subseteq \R^n$. For instance, if $S_\Phi$ consists of B\'ezier curves, $\Phi$ contains a list of control points. Given a target shape $T \subseteq \R^n$, we formulate fitting a parametric shape to approximate  $T$ w.r.t.\ $\Psi$ as minimizing
\begin{equation}\label{eq:generalloss}
    f_\Psi(\Phi) = \LL_\Psi[S_\Phi, T].
\end{equation}
For optimal shape parameters, $\hat\Phi := \argmin_\Phi f_\Psi(\Phi)$.  We propose two discrepancy measures, providing loss functions that capture different geometric features.

\subsection{Surface Loss}

We define surface discrepancy to be
\begin{equation}
\Psi^\textrm{surf}_{A,B}(x)\!=\!\delta\{\ker \dd^2_A\}(x)\!\dd^2_B(x)\!+\!\delta\{\ker \dd^2_B\}(x)\!\dd^2_A(x)
\end{equation}
where $\delta\{X\}$ is the Dirac delta defined uniformly on $X$, and $\ker f$ denotes the zero level-set of $f$.
$\Psi^\textrm{surf} > 0$ iff the shapes do not match, making it sensitive to local geometry:

\begin{prop}
The symmetric variational Chamfer distance between $A,B \subseteq \R^n$ is equal to the surface loss between $A$ and $B$, i.e., $\Ch^\mathrm{var}(A,B) = \LL_{\Psi^\textrm{surf}_{A,B}}$.
\end{prop}

\noindent Unlike Chamfer distance, the discrete version of our surface loss can be approximated efficiently without sampling points from either the parametric or target shape via evaluation over a regular grid, as we show in \S\ref{sec:final-loss}.

\subsection{Normal Alignment Loss}

We define normal alignment discrepancy to be
\begin{equation}
    \Psi^\textrm{align}_{A,B}(x) = 1 - \langle\nabla \dd_A(x), \nabla\dd_B(x)\rangle^2.
\end{equation}
Minimizing $f_{\Psi^\textrm{align}}$ aligns normals of the predicted primitives to those of the target. Following Figure~\ref{fig:chamfer-drawbacks}b, if $A$ contains dense vertical lines and $B$ contains horizontal lines, $\LL_{\Psi^\textrm{align}_{A,B}} \gg 0$ while $\Ch(A,B) \approx 0$.

\subsection{Final Loss Function}
\label{sec:final-loss}

\begin{figure}[t]
\centering
\includegraphics[width=\linewidth]{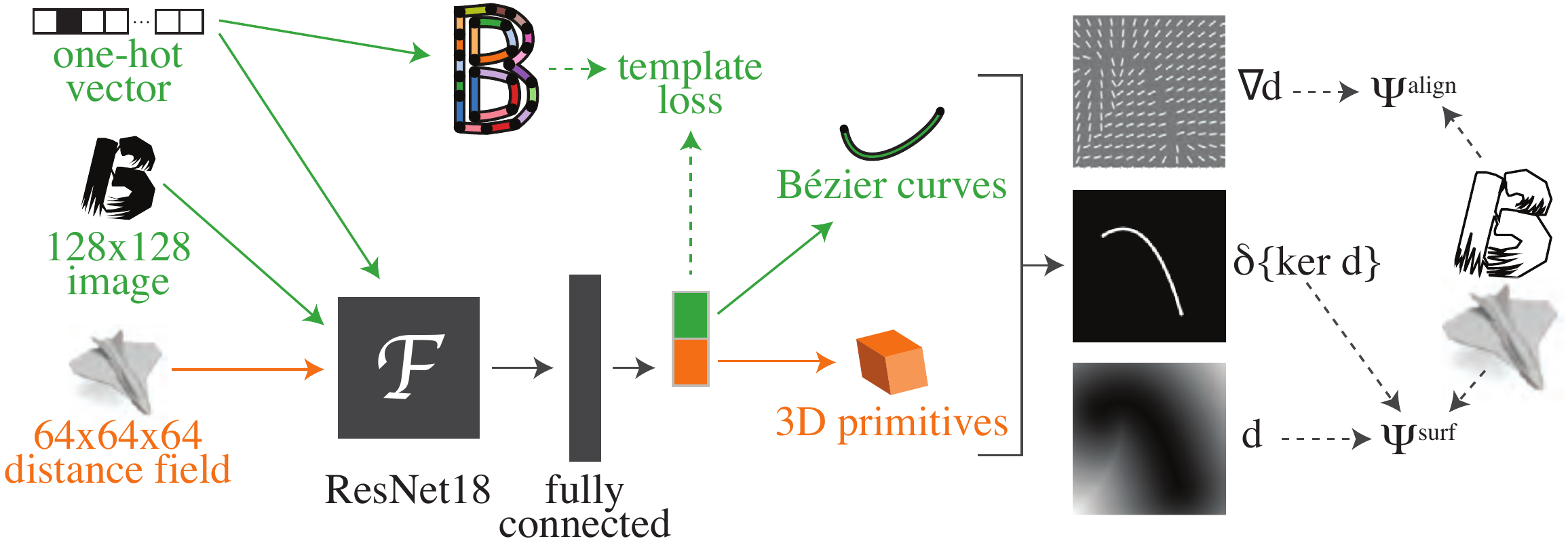}\vspace{-0.1in}
\caption{An overview of our pipelines---font vectorization (green) and 3D abstraction (orange).\vspace{-0.1in}}
\label{fig:architecture}
\end{figure}

The general distance field loss and proposed discrepancy measures are differentiable w.r.t.\ the shape parameters $\Phi$, as long as $\dd_{S_\Phi}$ is differentiable w.r.t.\ $\Phi$. Thus, they are well-suited to be optimized by a deep network predicting parametric shapes. We approximate \eqref{gdf-loss} via Monte Carlo integration:
\begin{equation}
	\LL_\Psi[A, B] \approx \frac{1}{|G|} \sum_{x \in G} \Psi_{A,B}(x),
\end{equation}
where $G$ is a 2D or 3D grid.

While we use a voxel grid grid to compute the integrals in our loss function, the resolution of the voxel grid only affects quadrature without limiting the resolution of our representation. The grid dictates how we sample distance values; the values themselves are derived from a continuous parametric representation. A small subvoxel change in the geometry will affect the distance value at multiple discrete voxels. This property is in distinct contrast to representations that only consider the occupancy grid of a shape---the resolution of such representations is strictly limited by the grid resolution.

For $\Psi^\textrm{surf}$, we use $\operatorname{Smootherstep}(1-\dd^2_A/\gamma^2)$ (with Smootherstep defined as in \cite{ebert2003texturing}) as a smooth version of $\delta\{\ker \dd^2_A\}$ to evaluate the expression on a grid and to avoid discontinuities, enabling smooth gradients in our optimization. We set $\gamma$ to twice the diameter of a voxel. For $\Psi^\textrm{align}$, we approximate gradients using finite differences.

We minimize $f_\Psi = f_{\Psi^\text{surf}}+\alpha^\textrm{align}f_{\Psi^\text{align}}$,
determining $\alpha^\textrm{align}=0.01$ for all experiments using cross-validation.

\subsection{Network Architecture and Training}
The network takes a $128\!\times\!128$ image or a $64\!\times\!64\!\times\!64$ distance field as input and outputs a parametric shape. We encode our input to a $\R^{256}$ latent space using a ResNet-18 \cite{he2016deep} architecture. We then use a fully connected layer with 256 units and ReLU nonlinearity followed by a fully connected layer with number of units equal to the dimension of the target parameterization. We pass the output through a sigmoid and rescale it depending on the parameters being predicted. Our pipeline is illustrated in Figure~\ref{fig:architecture}. We train each network on a single Tesla GeForce GTX Titan X GPU for approximately one day, using Adam \cite{kingma2014adam} with learning rate $10^{-4}$ and batch size 32 for 2D and 16 for 3D.

%% file: sections/2d.tex
\section{2D: Font Exploration and Manipulation}

We demonstrate our method in 2D for font glyph vectorization. Given a raster image of a glyph, our network outputs control points defining a collection of quadratic B\'ezier curves that approximate its outline. We produce nearly exact vector representations of glyphs from simple (non-decorative) fonts. From a decorative glyph with fine-grained detail, however, we recover a good approximation of the glyph's shape using a small number of B\'ezier primitives and a consistent structure. This process can be interpreted as projection onto a common latent space of control points.

We first describe our choice of primitives as well as the computation of their distance fields. We introduce a template-based approach to allow our network to better handle multimodal data (different letters) and test several applications.

\subsection{Approach}
\label{sec:approach}

\paragraph*{Primitives.}

We wish to use a 2D parametric shape primitive that is sparse and expressive and admits an analytic distance field. Our choice is the \emph{quadratic B\`ezier curve} (which we refer to as \emph{curve}), parameterized by control points $a, b, c \in \R^2$ and defined by $\gamma(t) = (1-t)^2a + 2(1-t)tb + t^2c,$ for $0 \le t \le 1$. We represent 2D shapes as the union of $n$ curves parameterized by $\Phi = \{a_i, b_i, c_i\}_{i=1}^n \subseteq \R^{3n}$.

\vspace{-0.05in}
\begin{prop}
Given a curve $\gamma$ parameterized by $a,b,c \in \R^2$ and a point $p \in \R^2$, the $\hat t \in \R$ such that $\gamma(\hat t)$ is the closest point on the curve to $p$ satisfies the following:
\begin{equation}
\label{curve-dist}
\begin{split}
\langle B, B \rangle \hat t^3 + 3 \langle A, B \rangle \hat t^2 &+ (2\langle A, A \rangle + \langle B, a - p \rangle)\hat t \\&+ \langle A, a - p \rangle = 0,
\end{split}
\end{equation}
where $A = b-a$ and $B = c - 2b + a$.
\end{prop}

Thus, evaluating the distance to a single curve $\dd_{\gamma_i}(p) = \|p - \gamma_i(\hat t)\|_2$ requires finding the roots of a cubic \cite{qin2006real}, which we can do analytically. To compute distance to the union of the curves, we take a minimum: $\dd_\Phi(p) = \min_{i=1}^n \dd_{\gamma_i}(p)$.

\begin{figure}[t]
\centering
\includegraphics[width=\linewidth]{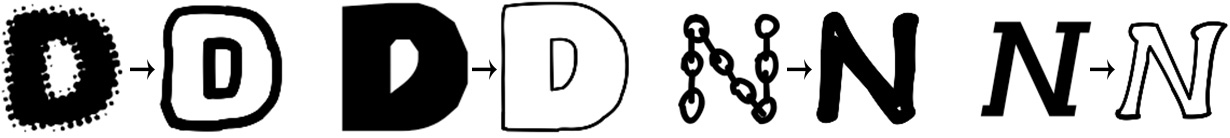}\vspace{-.1in}
\caption{Glyphs with predicted boundary curves rendered with predicted stroke thickness. The network thickens curves to account for stylistic details at the glyph boundaries.}
\label{fig:abc-stroke}
\end{figure}

In addition to the control points, we predict a stroke thickness for each curve. We use this parameter when computing the loss by ``lifting'' the predicted distance field, thus thickening the curve---if curve $\gamma$ has thickness $s$, we set $d^s_\gamma(p) = \min(d_\gamma(p)-s,0)$. While we do not visualize stroke thickness in our experiments, this approach allows the network to thicken curves to better match high-frequency filigree (see Figure~\ref{fig:abc-stroke}). This thickening is a simple operation in our distance field representation; sampling-based methods do not provide a natural way to thicken predicted geometry.

\vspace{-0.1in}
\paragraph*{Templates.}

Our training procedure is self-supervised, as we do not have ground truth curve annotations. To better handle the multimodal nature of our entire dataset with a single network, we label each training example with its letter, passed as additional input. This allows us to condition on input class by concatenating a 26-dimensional one-hot vector to the input, a common technique for conditioning \cite{zhu2017toward}.

We choose a ``standard'' curve representation per letter, capturing each letter's distinct geometric and topological features, by designing 26 templates from a shared set of control points. A \emph{template} of type $\ell \in \{\textrm{A}, \ldots, \textrm{Z}\}$ is a collection of points $T_\ell = \{p_1,\ldots,p_n\} \subseteq \R^{2n}$ with corresponding \emph{connectivity} determining how the points define curves. Since our curves form closed loops, we reuse endpoints.

\begin{figure}[t]
\centering
\includegraphics[width=\linewidth]{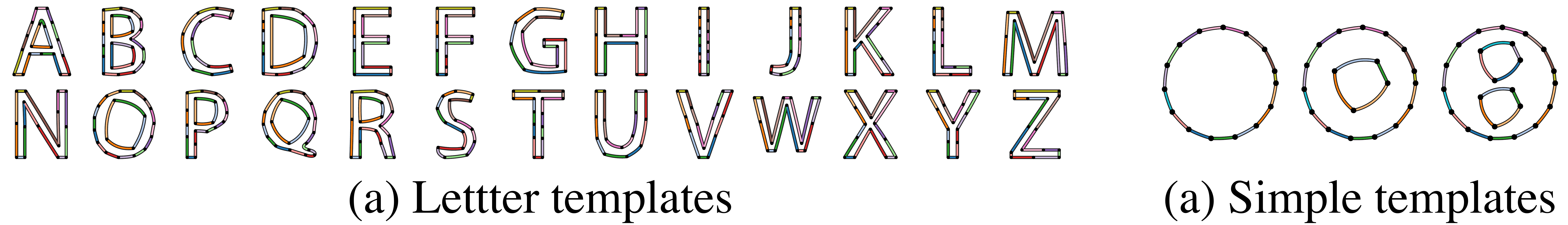}\vspace{-.1in}
\caption{Font glyph templates. These determine the connectivity and initialize the placement of the predicted curves.\vspace{-0.1in}}
\label{fig:templates}
\end{figure}

For glyph boundaries of uppercase English letters, there are three connectivity types---one loop (e.g., ``C"), two loops (e.g., ``A"), and three loops (``B"). In our templates, the first loop has 15 curves and the other loops have 4 curves each. We will show that while letter templates (Figure~\ref{fig:templates}a) better specialize to the boundaries of each glyph, we still achieve good results with simple templates (Figure~\ref{fig:templates}b). Even without letter-specific templates, our system learns a consistent geometric representation, establishing cross-glyph correspondences purely using self-supervision.

We use predefined templates together with our labeling of each training example for two purposes. First, connectivity is used to compute curve control points from the network output. Second, they provide a \emph{template loss}:
\begin{equation}
\LL^\textrm{template}(\ell, x) = \alpha^\textrm{template}e^{(t/s)} \| T_\ell -  h^t(x)\|_2^2,
\end{equation}
where $s \in \Z_+$,  $\gamma \in (0,1)$, $t$ is the iteration number, $x$ is the input image, and $h^t(x)$ is the network output at iteration $t$. This initializes the network output, such that an input of type $\ell$ initially maps to template $\ell$. As this term decays, the other loss terms take over. We set $\alpha^\textrm{template}=10$ and $s=500$, though other choices of parameters for which the template term initially overpowers the rest of the loss also work.

\subsection{Experiments}

We train our network on the 26 uppercase English letters extracted from nearly 10,000 fonts. The input is a raster image of a letter, and the target distance field to the boundary of the original vector representation is precomputed.

\vspace{-0.1in}
\paragraph*{Ablation study.}

We demonstrate the benefit of our loss over Chamfer distance as well as the contribution of each of our loss terms. While having 26 unique templates helps achieve better results, it is not crucial---we evaluate a network trained with  three ``simple templates" (Figure~\ref{fig:templates}b), which capture the three topology classes of our data.

\begin{table}[ht]
    \centering
    \begin{tabular}{|c|c|}
         \hline
         Model & Average error \\
         \hline\hline
         \bf Full model (ours) & \bf 0.509 \\
         No surface term (ours) & 1.613 \\
         No alignment term (ours)& 0.642 \\
         Simple templates (ours)  & 0.641 \\
         Chamfer (with letter templates) & 0.623 \\
         AtlasNet \cite{groueix2018atlasnet} & 5.154 \\
         \hline
    \end{tabular}
    \caption{Comparison between subsets of our full model as well as standard Chamfer distance and AtlasNet. Average error is Chamfer distance (in pixels on a $128\!\times\!128$ image) between ground truth and uniformly sampled predicted curves.\vspace{-0.1in}}
    \label{table:chamfer}
\end{table}

For the Chamfer loss experiment, we use the same hyperparameters as for our method and sample 5,000 points from the source and target geometry. We initialize the model output to the full letter templates, like in our full model.

We also evaluate on 20 sans-serif fonts, computing Chamfer distance between our predicted curves and ground truth geometry, sampling uniformly (average error in Table~\ref{table:chamfer}). Uniform sampling is a computationally-expensive and non-differentiable procedure only for evaluation \emph{a posteriori}---not suitable for training. While it does not correct all of Chamfer distance's shortcomings, we use it as a baseline to evaluate quality. We limit to sans-serif fonts since we do not expect to faithfully recover local geometry. Our full loss outperforms Chamfer loss, and both our loss terms are necessary. Figure~\ref{fig:abc-ablation} shows qualitative results on test set glyphs; see supplementary material for additional results.

\begin{figure}[ht]
    \centering
    \includegraphics[width=0.6\linewidth]{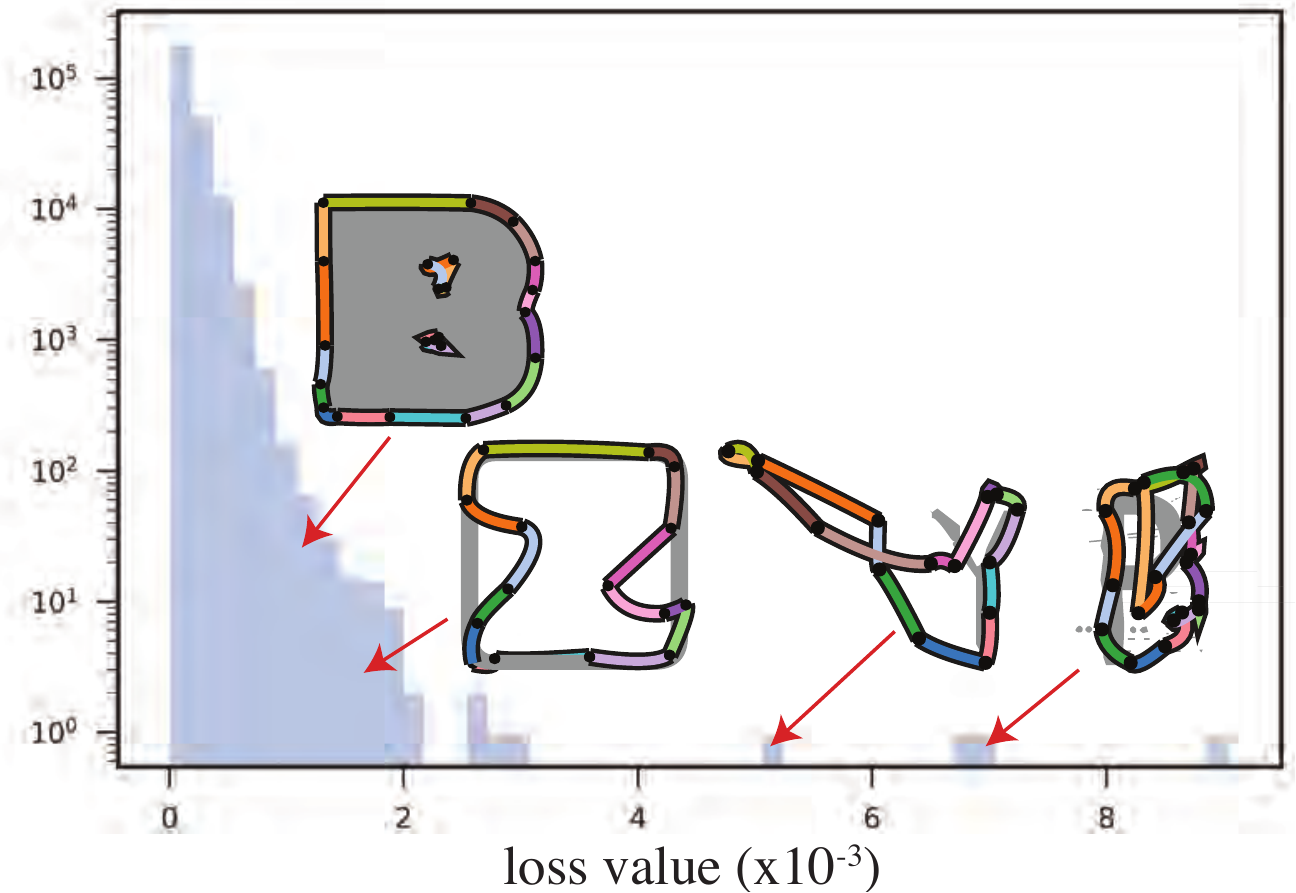}
    \caption{Number of examples per quantized loss value. We visualize the input and predicted curves for several outliers.\vspace{-0.2in}}
    \label{fig:abc-loss}
\end{figure}

We demonstrate robustness in Figure~\ref{fig:abc-loss} by quantizing our loss values and plotting the number of examples for each value. High loss outliers are generally caused by noisy data---they are either not uppercase English letters or have fundamentally uncommon structure.

\begin{figure}[t]
    \centering
    \includegraphics[width=\linewidth]{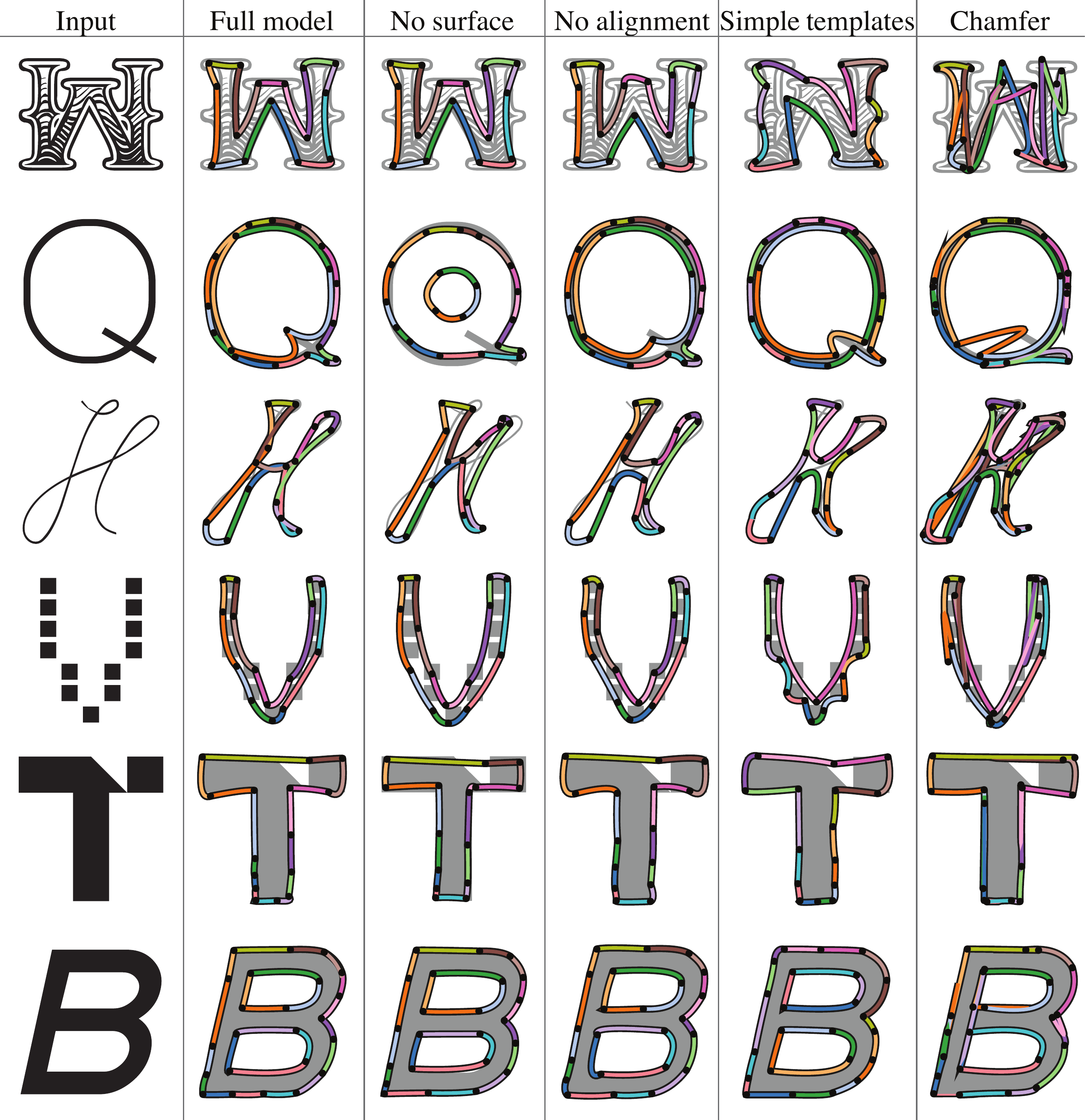}\vspace{-.1in}
    \caption{Ablation study and comparison to Chamfer.\vspace{-0.1in}}
    \label{fig:abc-ablation}
\end{figure}

\begin{figure*}[t]
    \centering
    \begin{subfigure}[b]{.4\linewidth}
        \centering
        \includegraphics[width=\linewidth]{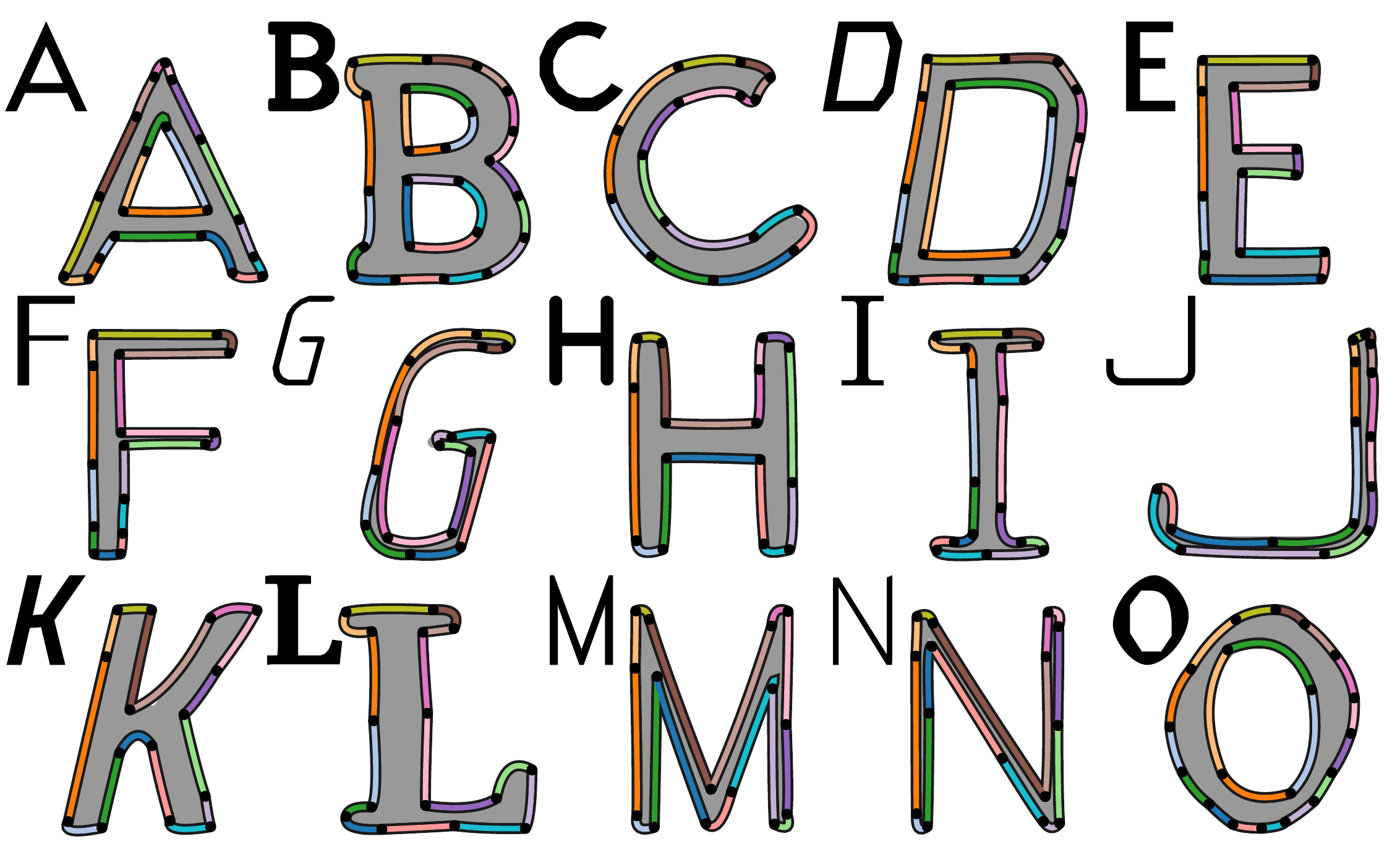}
        \caption{Plain font glyphs}
        \label{fig:abc-plain}
    \end{subfigure}
    \hspace{8mm}
    \begin{subfigure}[b]{.4\linewidth}
        \includegraphics[width=\linewidth]{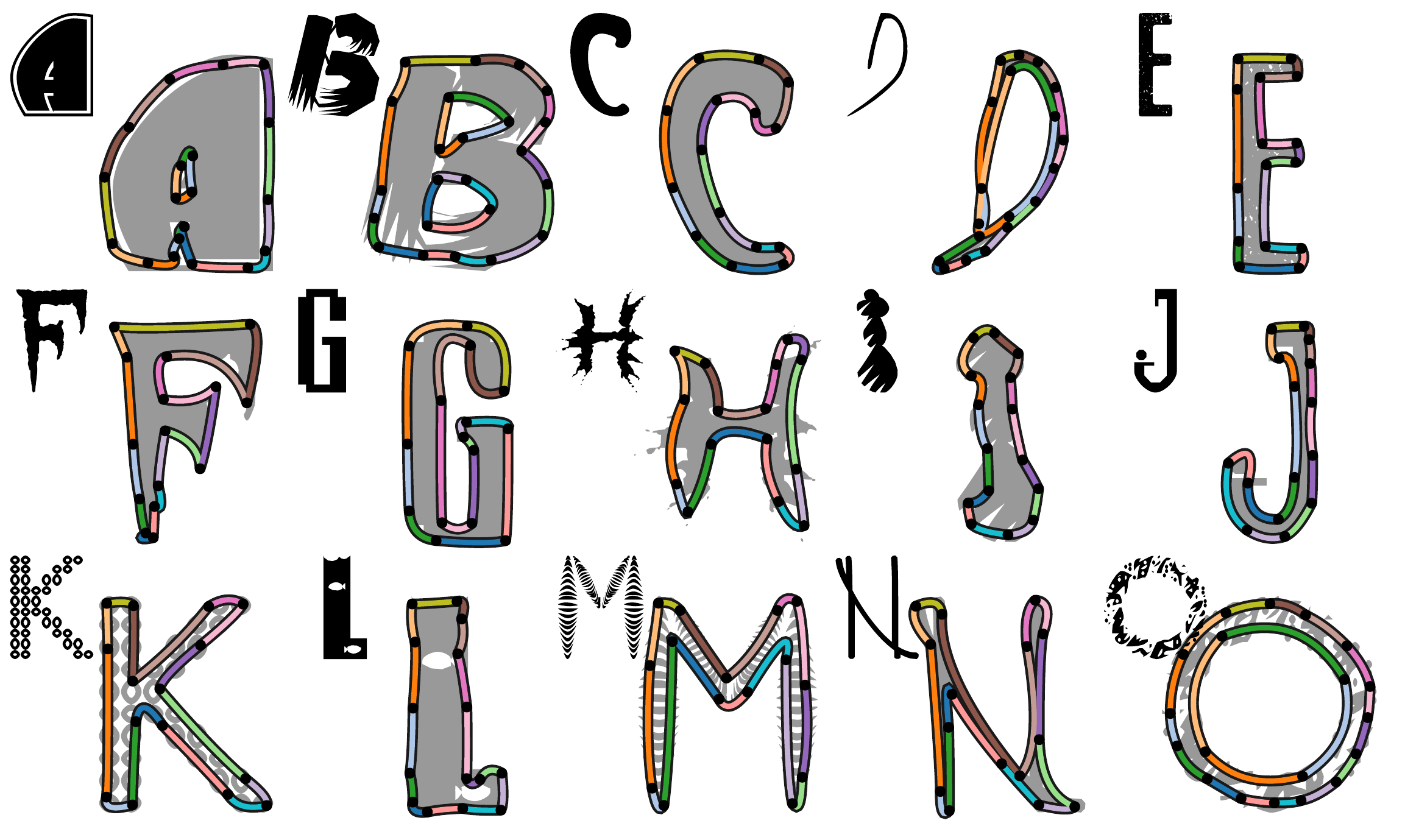}
        \caption{Decorative font glyphs}
        \label{fig:abc-decorative}
    \end{subfigure}\vspace{-.12in}
    \caption{Vectorization of various glyphs. For each we show the raster input (top left,black) along with the vectorization (colored curves) superimposed. When the input has simple structure (a), we recover an accurate vectorization. For fonts with decorative details (b), our method places curves to capture overall structure. Results are taken from the test dataset.\vspace{-.18in}}
\end{figure*}

\vspace{-0.1in}
\paragraph*{Comparison to AtlasNet.}

In AtlasNet \cite{groueix2018atlasnet}, geometry is reconstructed by training implicit decoders, which map a point in the unit square to a point on the target surface, optimizing Chamfer distance. We modify the AtlasNet system to our task and demonstrate that our method method proposes a more effective geometry representation and loss.

AtlasNet represents shapes as points in a learned high dimensional space, which does not obviously correlate to geometric features. Thus, in contrast to our explicit representation as a collection of control points, it does not facilitate geometric interpretability. Additionally, this makes it difficult to impose geometric priors---it is unclear how to initialize AtlasNet to predefined templates, as we do in \S\ref{sec:approach}.

For a fair comparison, we train an AtlasNet model that maps points from the boundary of a circle (rather than the interior of a square) into 2D. We only train on letters with single loop topology (\emph{C, E, F}, etc.) and sample 5,000 points. Thus, this setting is comparable to the \emph{simple templates} experiment from our ablation.

\begin{figure}[t]
    \centering
    \includegraphics[width=\linewidth]{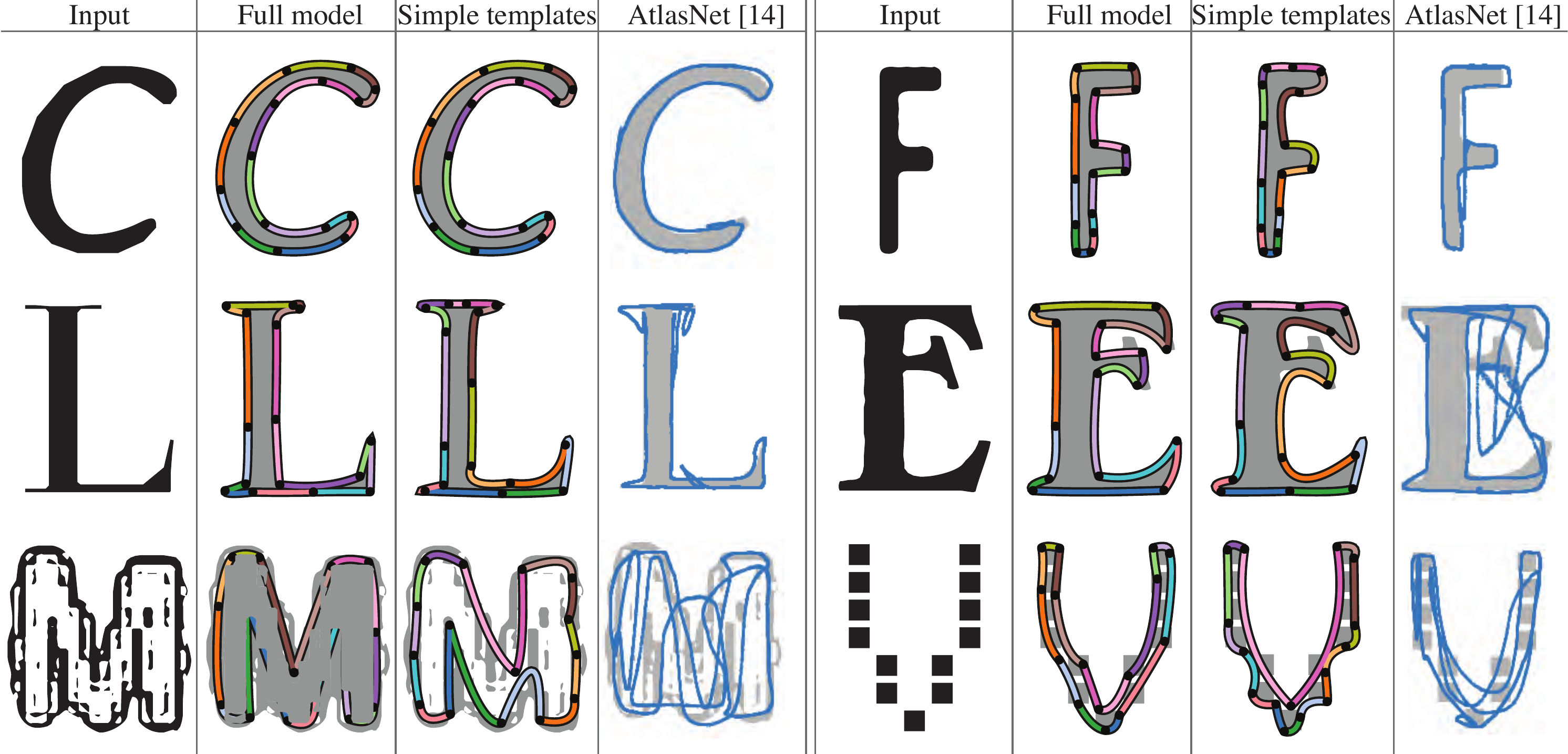}\vspace{-.1in}
    \caption{Comparison to AtlasNet \cite{groueix2018atlasnet} with a closed loop start shape to our simple templates and full models. We only train (and test) AtlasNet on letters with a single loop.\vspace{-0.1in}}
    \label{fig:abc-atlasnet}
\end{figure}

We show results in Figure~\ref{fig:abc-atlasnet}. Although AtlasNet recovers overall structure of the input, it suffers from artifacts, self-intersections, and imprecision not exhibited by our method, even with simple templates. Likely, this is due to the fact that AtlasNet exhibits the drawbacks of Chamfer distance identified in \S\ref{sec-prelim}, i.e., non-uniform sampling and lack of sensitivity to normal alignment. We include a quantitative comparison in Table~\ref{table:chamfer}. Our method outperforms AtlasNet even based on on a uniformly-sampled Chamfer distance metric.

\vspace{-0.1in}
\paragraph*{Vectorization.}
For any font glyph, our method generates a consistent sparse vector representation, robustly and accurately describing the glyph's structure while ignoring decorative and noisy details. For simple fonts, our representation is a near-perfect vectorization, as in Figure~\ref{fig:abc-plain}. For decorative glyphs, our method produces a meaningful abstraction. While a true vectorization would contain many curves with a large number of connected components, we succinctly capture the glyph's overall structure (Figure~\ref{fig:abc-decorative}).

Our method preserves semantic correspondences. The same curve is consistently used for the boundary of, e.g., the top of an ``I". These correspondences persist \emph{across} letters with both full and simple templates---see, e.g., the ``E" and ``F" in Figure~\ref{fig:abc-plain} and \ref{fig:abc-decorative} and ``simple templates'' in Figure~\ref{fig:abc-ablation}.

\vspace{-0.1in}
\paragraph*{Retrieval and exploration.}
Our sparse representation can be used to explore the space of glyphs, useful for artists and designers, without the need for manual labelling or annotation. Treating control points as a metric space, we can perform Euclidean nearest-neighbor lookups for font retrieval.

\begin{figure}[ht]
    \centering
    \includegraphics[width=0.8\linewidth]{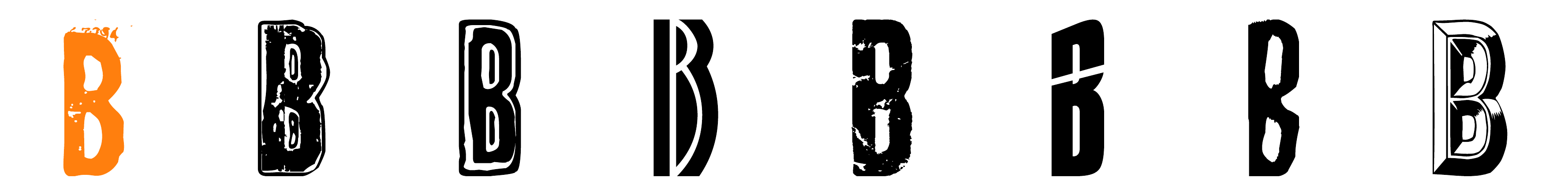}
    \includegraphics[width=0.8\linewidth]{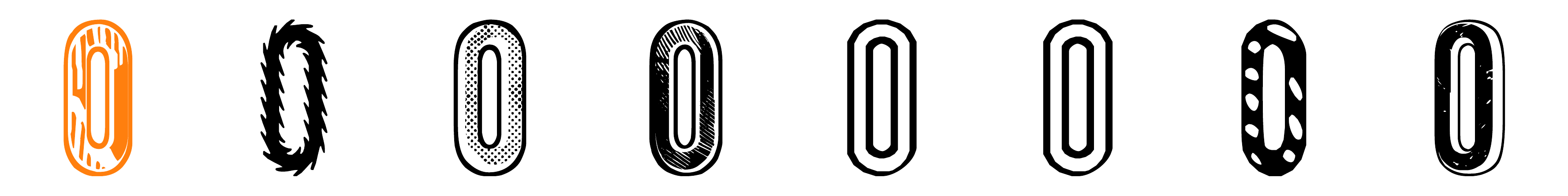}
    \vspace{-0.1in}
    \caption{Nearest neighbors for a glyph in curve space, sorted by proximity. The query glyph is in orange.\vspace{-0.05in}}
    \label{fig:abc-nn}
\end{figure}

In Figure~\ref{fig:abc-nn}, for each query glyph, we compute its curve representation and retrieve seven nearest neighbors in curve space. Because our representation captures geometric structure, we find glyphs that are similar structurally, despite decorative and stylistic differences.

\begin{figure}[ht]
    \centering
    \includegraphics[width=0.8\linewidth]{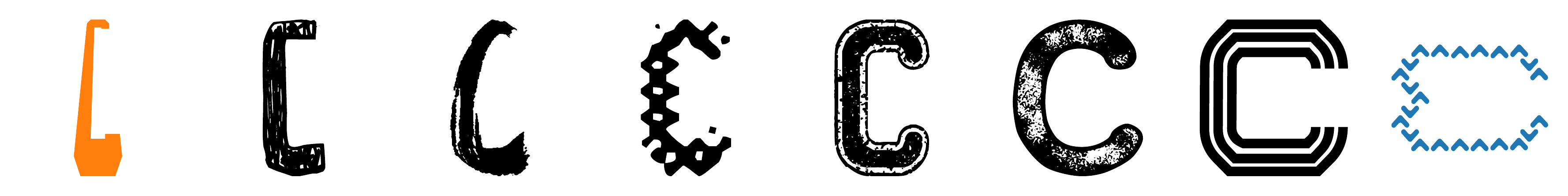}
    \includegraphics[width=0.8\linewidth]{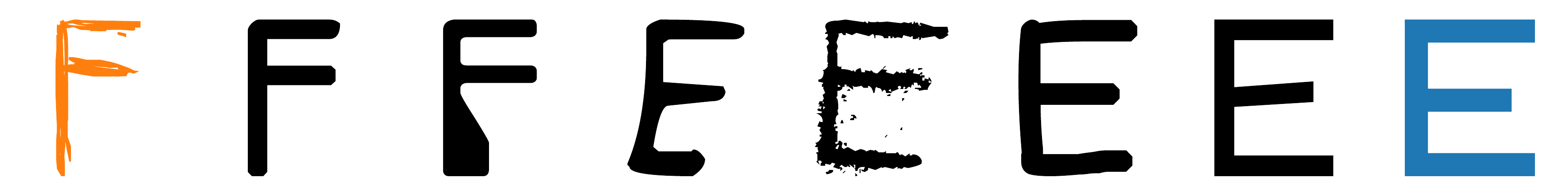}
    \vspace{-0.1in}
    \caption{Interpolating between fonts in curve space. The start and end are in orange and blue, respectively, and the nearest glyphs to linear interpolants are shown in order.\vspace{-0.05in}}
    \label{fig:abc-interpolation} 
\end{figure}

We can also consider a path in curve space starting at the curves for one glyph and ending at those for another. By sampling nearest neighbors along this trajectory, we ``interpolate" between glyphs. As in Figure~\ref{fig:abc-interpolation}, this produces meaningful collections of fonts for the same letter and reasonable results when the start and end glyphs are different letters. Additional results are in supplementary material.

\begin{figure}[ht]
    \centering
    \includegraphics[width=0.6\linewidth]{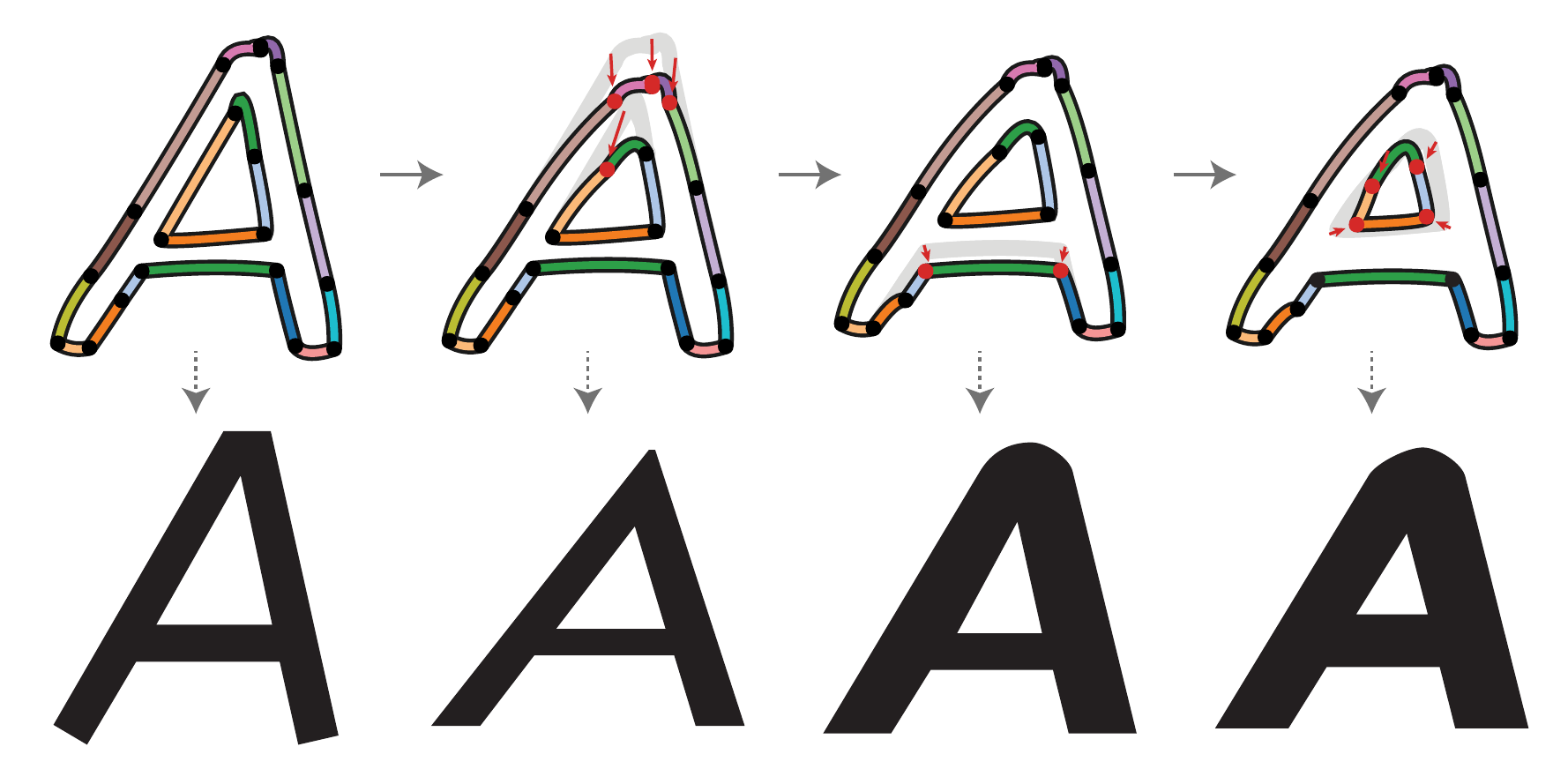}\vspace{-.1in}
    \caption{User-guided font exploration. At each edit, the nearest glyph is displayed below. This lets the user explore the dataset through geometric refinements.\vspace{-0.05in}}
    \label{fig:abc-exploration}
\end{figure}

Nearest-neighbor lookups in curve space also can help find a font matching desired geometric characteristics. A possible workflow is in Figure~\ref{fig:abc-exploration}---through incremental refinements of the curves the user can quickly find a font.

\begin{figure}[t]
    \centering
    \includegraphics[width=0.5\linewidth]{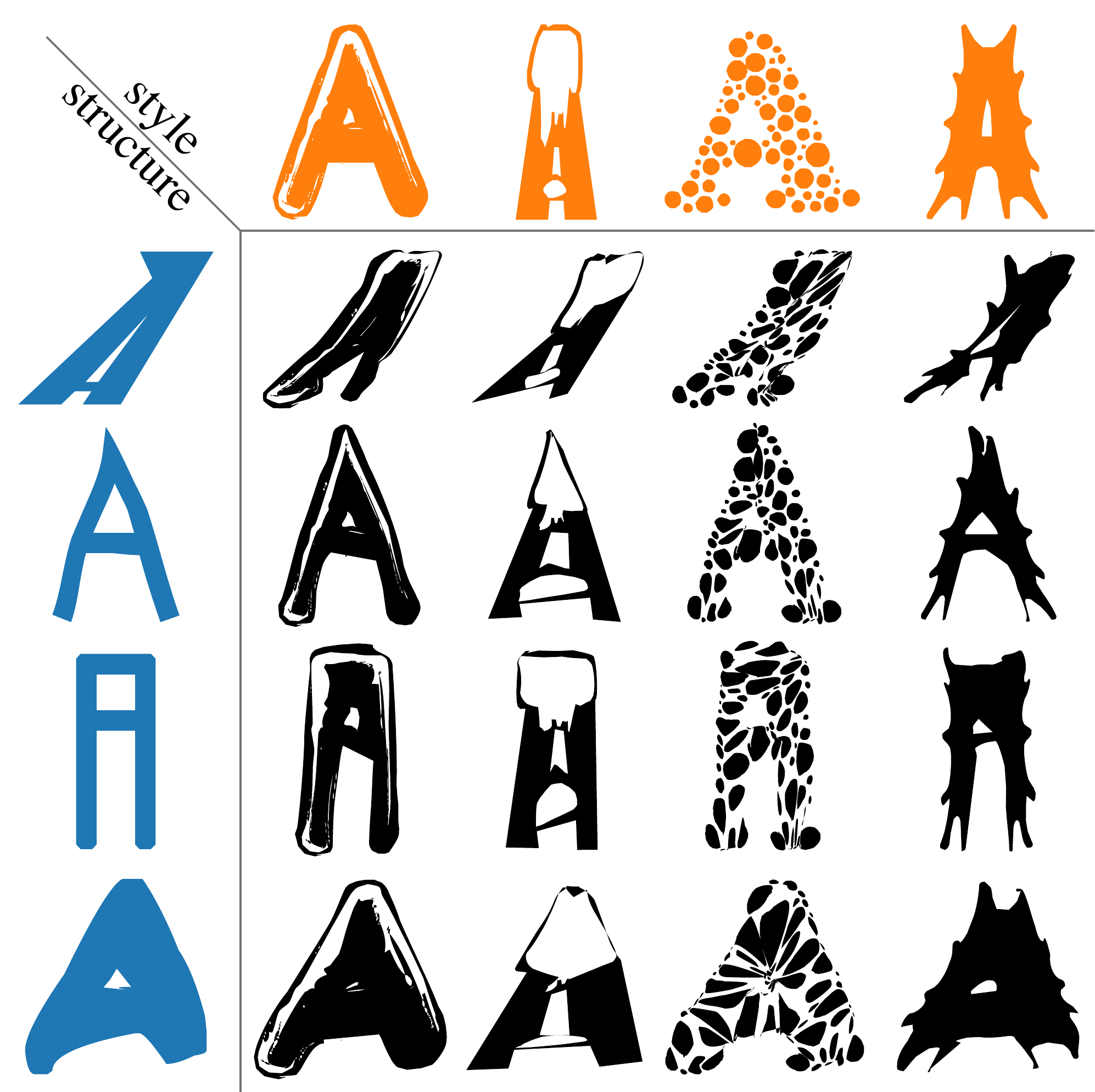}\vspace{-.1in}
    \caption{Mixing of style (columns) and structure (rows) of the \emph{A} glyph from different fonts. We deform each starting glyph (orange) into the structure of each target glyph (blue).\vspace{-0.1in}}
    \label{fig:abc-analogies}
\end{figure}

\vspace{-0.1in}
\paragraph*{Style and structure mixing.}
Our sparse curve representation describes geometric structure, ignoring stylistic and decorative details. We leverage this to warp a glyph with desired style to the structure of another glyph (Figure~\ref{fig:abc-analogies}).

We first generate the sparse curve representation for source and target glyphs. Since our representation uses the same set of curves, we can estimate dense correspondences and use them to warp original vectors of the source glyph to conform to the shape of the target. For each point on the source, we apply a translation that is a weighted sum of the translations from the sparse curve control points in the source glyph to those in the target glyph.

\begin{figure}[t]
    \centering
    \includegraphics[width=\linewidth]{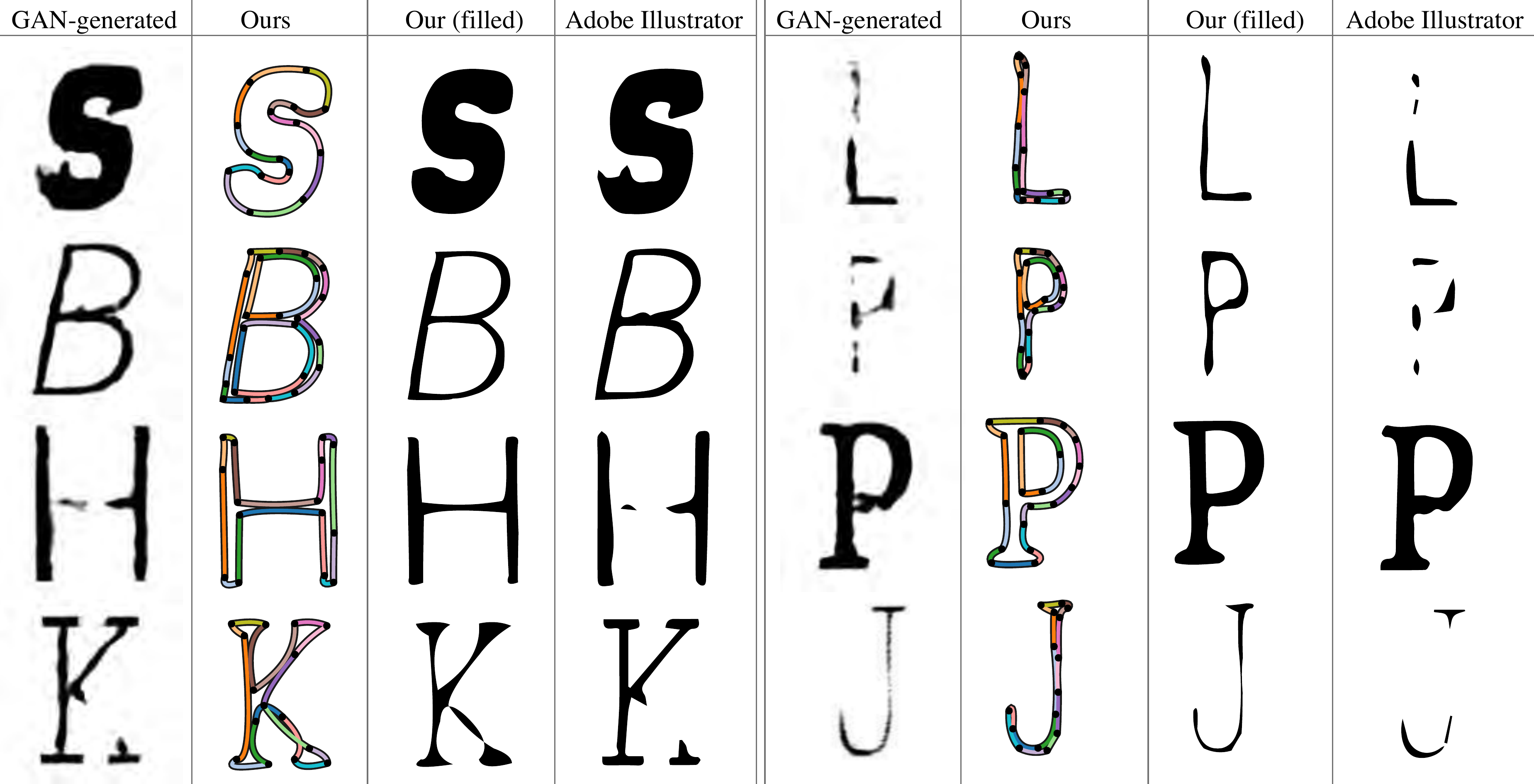}\vspace{-.1in}
    \caption{Vectorized GAN-generated fonts from \cite{azadi2018multi}.\vspace{-0.1in}}
    \label{fig:abc-gan}
\end{figure}

\vspace{-0.1in}
\paragraph*{Repair.}
Our system learns a strong prior on glyph shape, allowing us to robustly handle noisy input. In \cite{azadi2018multi}, a generative adversarial network (GAN) generates novel glyphs. The outputs, however, are raster images, often with noise and missing parts. Figure~\ref{fig:abc-gan} shows how our method can simultaneously vectorize and repair GAN-generated glyphs. Compared to a vectorization tool like Adobe Illustrator Live Trace, we infer missing data based on learned priors, making the glyphs usable starting points for font design.

\begin{figure}[t]
    \centering
    \includegraphics[width=\linewidth]{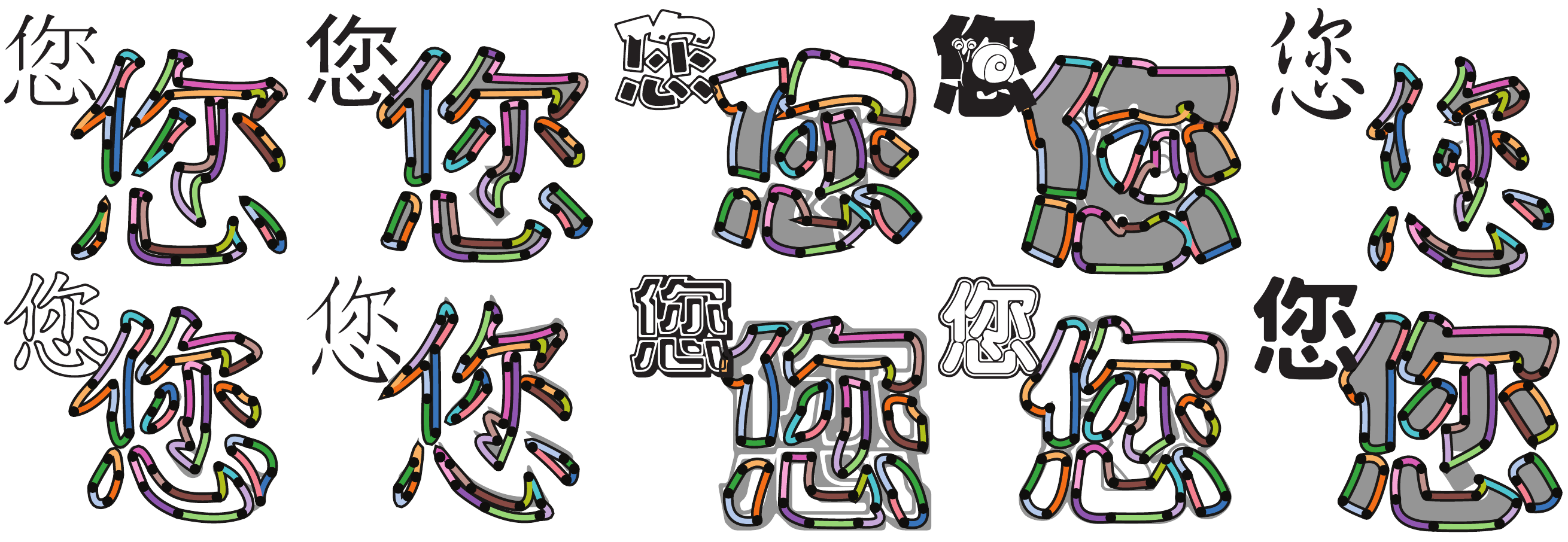}\vspace{-.1in}
    \caption{Vectorization of Chinese character \emph{您}.\vspace{-0.1in}}
    \label{fig:abc-chinese}
\end{figure}

\begin{CJK*}{UTF8}{gbsn}
\vspace{-0.1in}
\paragraph*{Other glyphs.}
Our method generalizes to more complex input than uppercase English glyphs. We demonstrate this by training a model to vectorize the Chinese character \emph{您}, which has significant geometric and topological complexity. We use a template that roughly captures the structure of the character. Results on several fonts are shown in Figure~\ref{fig:abc-chinese}.

\end{CJK*}

%% file: sections/3d.tex
\section{3D: Volumetric Primitive Prediction}

We reconstruct 3D surfaces out of various primitives, which allow our model to be expressive, sparse, and abstract.

\subsection{Approach}

Our first primitive is a \emph{cuboid}, parameterized by $\{b, t, q\}$, where $b = (w,h,d), t \in \R^3$ and $q \in \mathbb{S}^4$ a quaternion, i.e., an origin-centered (hollow) rectangular prism with dimensions $2b$ to which we apply rotation $q$ and then translation $t$.
\begin{prop}
	Let $C$ be a cuboid with parameters $\{b, t, q\}$ and $p \in \R^3$ a point. Then, the signed distance between $p$ and $C$ is
	\begin{equation}
		\dd_C(p) = \|\!\max(d,0)\!\|_2 + \min(\max(d_x, d_y, d_z),0),
	\end{equation}
	where $p'=q^{-1}(p-t)q$ using the Hamilton product and $d = (|p'_x|, |p'_y|, |p'_z|)-b$.
\end{prop}
Inspired by \cite{Paschalidou2019CVPR}, we additionally use a \emph{rounded cuboid} primitive by introducing a radius parameter $r$ and computing the signed distance by $\dd_{\mathit{RC}}(p)=\dd_C(p)-r$.

A unique advantage of our distance field representation is the ability to perform CSG boolean operations.  Since our distances are \emph{signed}, we can compute the distance to the union of $n$ primitives by taking a minimum over distance fields. With sampling-based methods such as Chamfer distance optimization, care must be taken to avoid sampling interior faces that are not part of the outer surface.

\subsection{Experiments}

We train on the airplane and chair categories of ShapeNet Core V2 \cite{shapenet2015}, taking as input a distance field. Thus, our method is fully self-supervised.

\vspace{-0.15in}
\paragraph*{Surface abstraction.}

\begin{figure}[t]
    \centering
    \includegraphics[width=\linewidth]{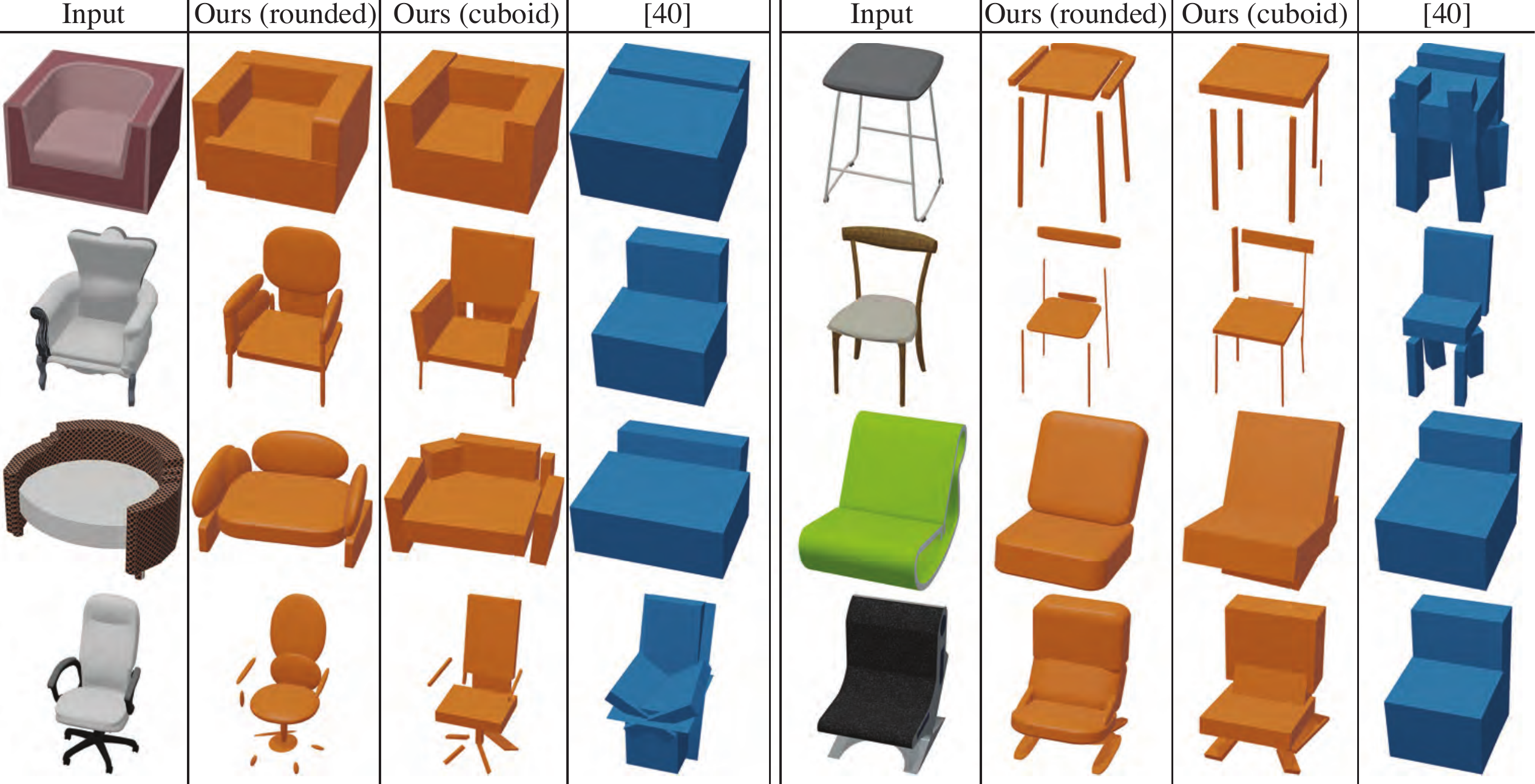}\vspace{-0.1in}
    \caption{Abstractions of test set chairs using our method and the method of \cite{tulsiani2017learning}.\vspace{-0.15in}}
    \label{fig:cube-chairs}
\end{figure}

\begin{figure}[t]
    \centering
    \includegraphics[width=0.85\linewidth]{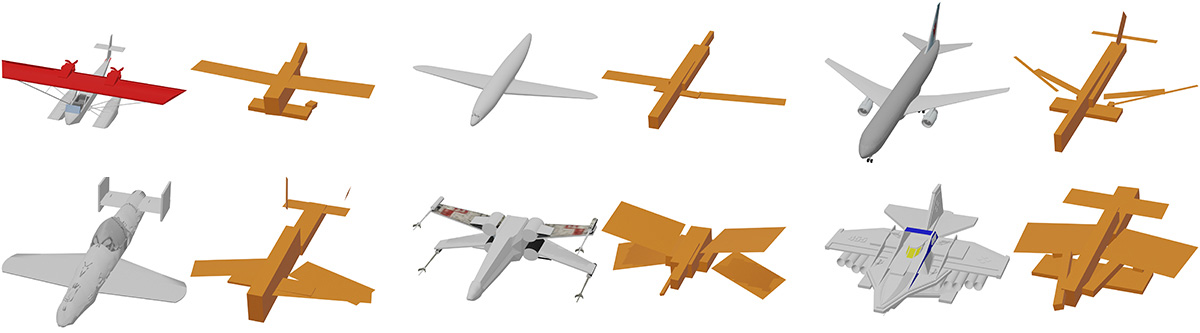}\vspace{-0.1in}
    \caption{Cuboid abstractions of test set airplanes.\vspace{-0.15in}}
    \label{fig:cube-airplanes}
\end{figure}

In Figure~\ref{fig:cube-chairs}, for each ShapeNet chair, we show the our cuboid abstraction, our rounded cuboid abstraction, and the abstraction of \cite{tulsiani2017learning}. We show our cuboid abstractions of ShapeNet airplanes in Figure~\ref{fig:cube-airplanes}. Each of our networks outputs 16 primitives, and we discard cuboids with high overlap using the method of \cite{tulsiani2017learning}. The resulting abstractions capture high-level structures of the input. See supplementary material for additional results.

\vspace{-0.15in}
\paragraph*{Segmentation.}

Because we place cuboids consistently, we can use them  for segmentation. Following \cite{tulsiani2017learning}, we demonstrate on the COSEG chair dataset. We first label each cuboid predicted by our network (trained on ShapeNet chairs) with a segmentation class (seat, back, legs). Then, we generate a cuboid decomposition of each chair in the dataset and segment according to the nearest cuboid. We achieve a mean accuracy of 94.6\%, exceeding the 89.0\% accuracy of \cite{tulsiani2017learning}.

\vspace{-0.15in}
\paragraph*{CSG operations.}
  
\begin{figure}[t]
    \centering
    \includegraphics[width=0.85\linewidth]{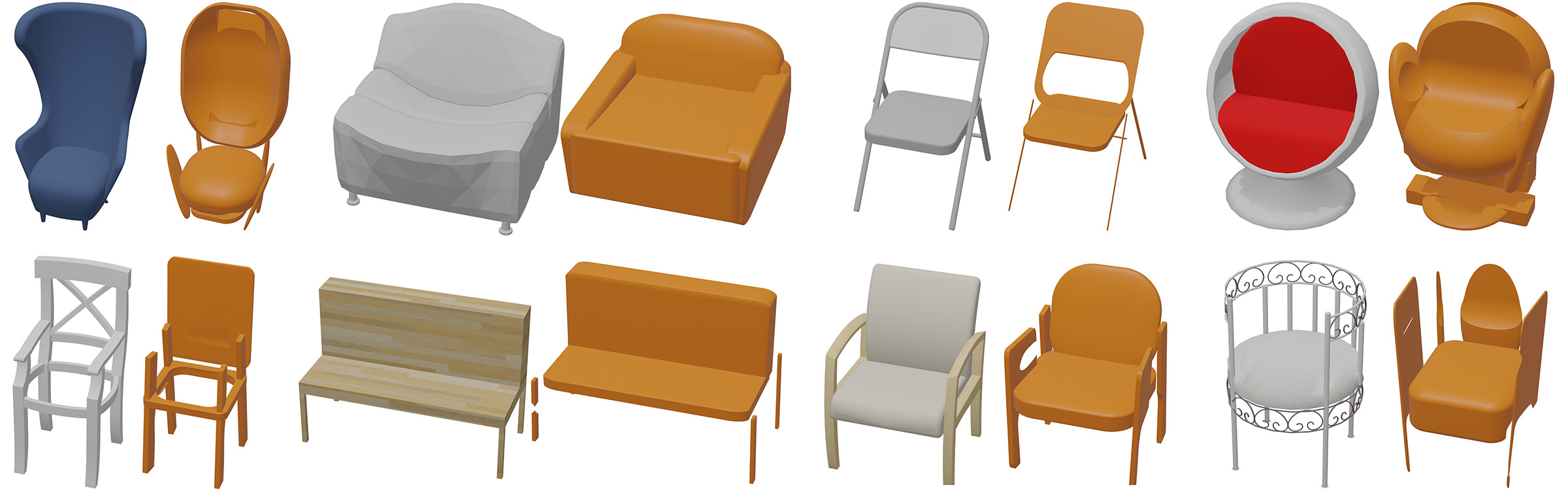}\vspace{-0.1in}
    \caption{Test set chair CSG abstractions. We predict eight rounded cuboids minus eight other rounded cuboids.\vspace{-0.2in}}
    \label{fig:csg}
\end{figure}

In Figure~\ref{fig:csg}, we show results of a network that outputs parameters for the union of eight rounded cuboids minus eight rounded cuboids. For inputs compatible with this template, we get good results. It is unclear how to achieve unsupervised CSG predictions using Chamfer loss.

%% file: sections/conclusion.tex
\section{Conclusion}

Representation is a key theme in deep learning---and machine learning more broadly---applied to geometry.  Assorted means of communicating a shape to and from a deep network present varying tradeoffs between efficiency, quality, and applicability.  While considerable effort has been put into choosing representations for certain tasks, the tasks we consider have \emph{fixed} representations for the input and output:  They take in a shape as a function on a grid and output a sparse set of parameters.  Using distance fields and derived functions as intermediate representations is natural and effective, not only performing well empirically but also providing a simple way to describe geometric loss functions.

Our learning procedure is applicable to many additional tasks.  A natural next step is to incorporate our network into more complex pipelines for tasks like vectorization of complex drawings~\cite{bessmeltsev2019vectorization}, for which the output of a learning procedure needs to be combined with classical techniques to ensure smooth, topologically valid output.  A challenging direction might be to incorporate user guidance into training or evaluation, developing the algorithm as a partner in shape reconstruction rather than generating a deterministic output.

Our experiments suggest several extensions for future work.  The key drawback of our approach is the requirement of closed-form distances for the primitives.  While there are many primitives that could be incorporated this way, a fruitful direction might be to alleviate this requirement, e.g. by including flexible implicit primitives like metaballs~\cite{blinn1982generalization}.  We could also incorporate more boolean operations into our pipeline, which easily supports them using algebraic operations on signed distances, in analogy to the CAD pipeline, to generate complex topologies and geometries with few primitives.  The combinatorial problem of determining the best sequence of boolean operations for a given input would be particularly challenging even for clean data~\cite{du2018inversecsg}.  Finally, it may be possible to incorporate our network into \emph{generative} algorithms to create new unseen shapes.

%% file: sections/supplementary.tex
\begin{figure*}[t]
    \centering
    \includegraphics[width=0.8\linewidth]{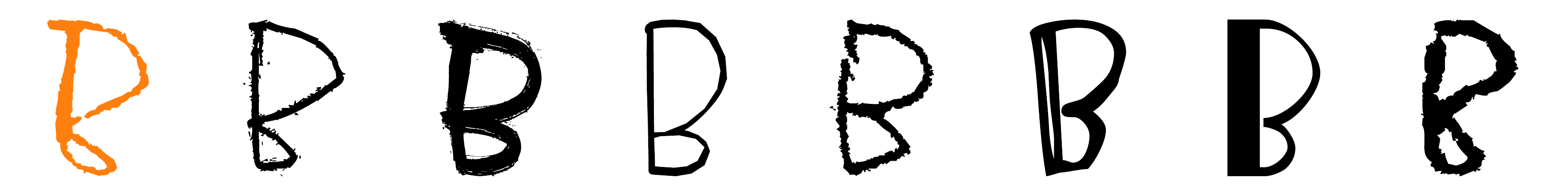}
    \includegraphics[width=0.8\linewidth]{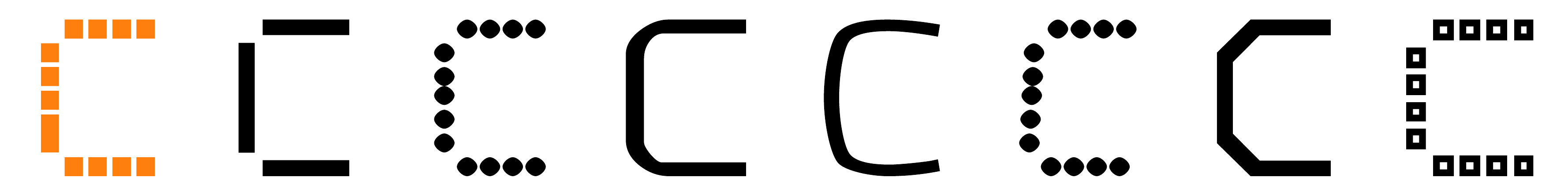}
    \includegraphics[width=0.8\linewidth]{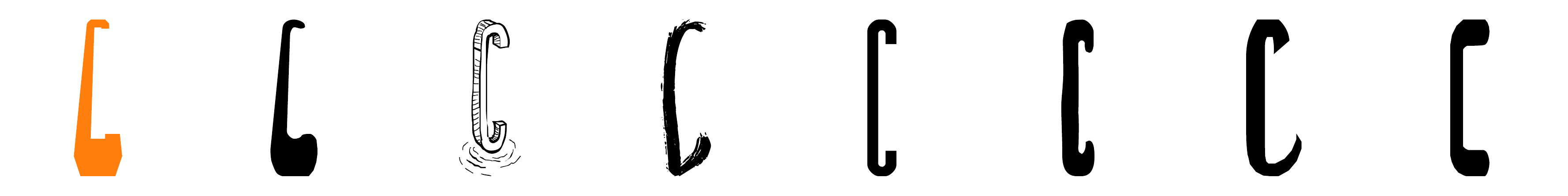}
    \includegraphics[width=0.8\linewidth]{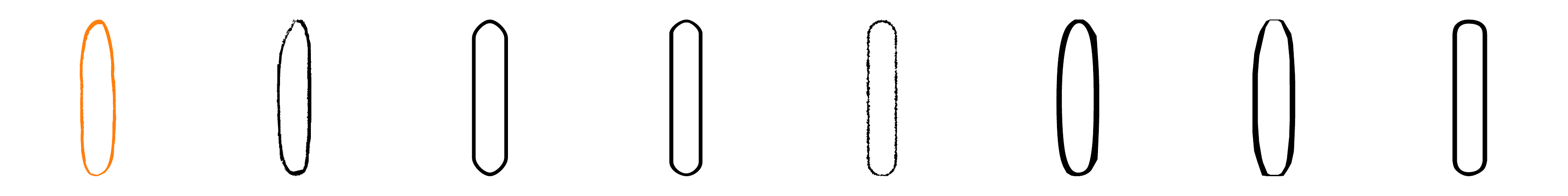}
    \caption{Glyph nearest neighbors in curve space.}
    \label{fig:sup-nn}
\end{figure*}
\begin{figure*}[t]
    \centering
    \includegraphics[width=0.8\linewidth]{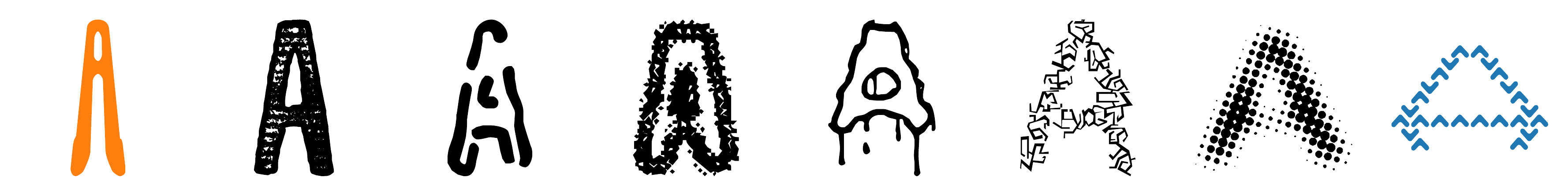}
    \includegraphics[width=0.8\linewidth]{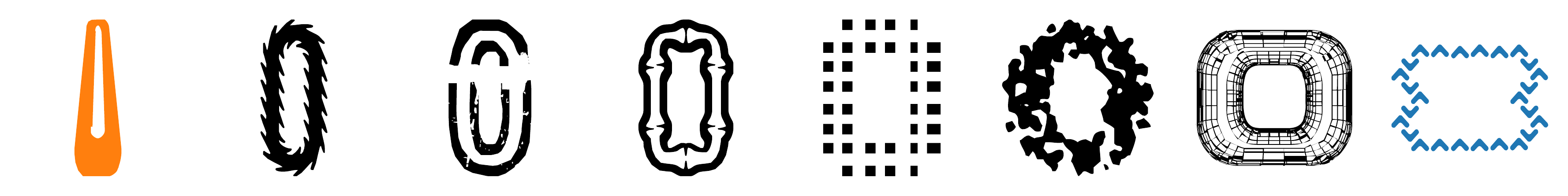}
    \includegraphics[width=0.8\linewidth]{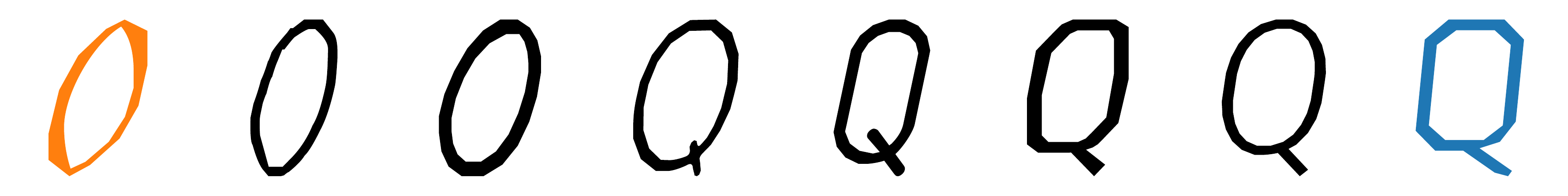}
    \includegraphics[width=0.8\linewidth]{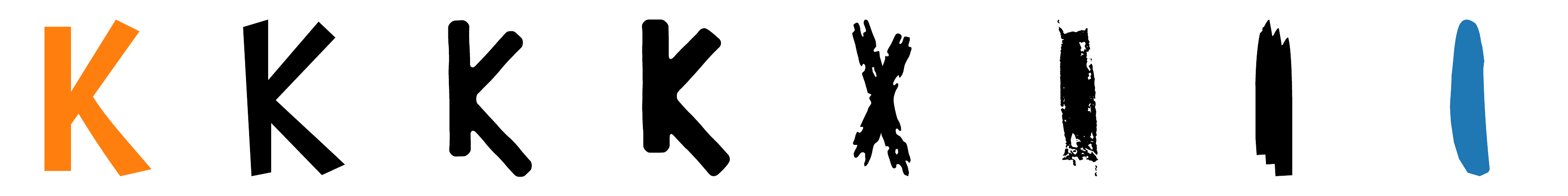}
    \caption{Interpolations between fonts in curve space.}
    \label{fig:sup-interpolation} 
\end{figure*}
\begin{figure*}[t]
    \centering
    \includegraphics[width=\linewidth]{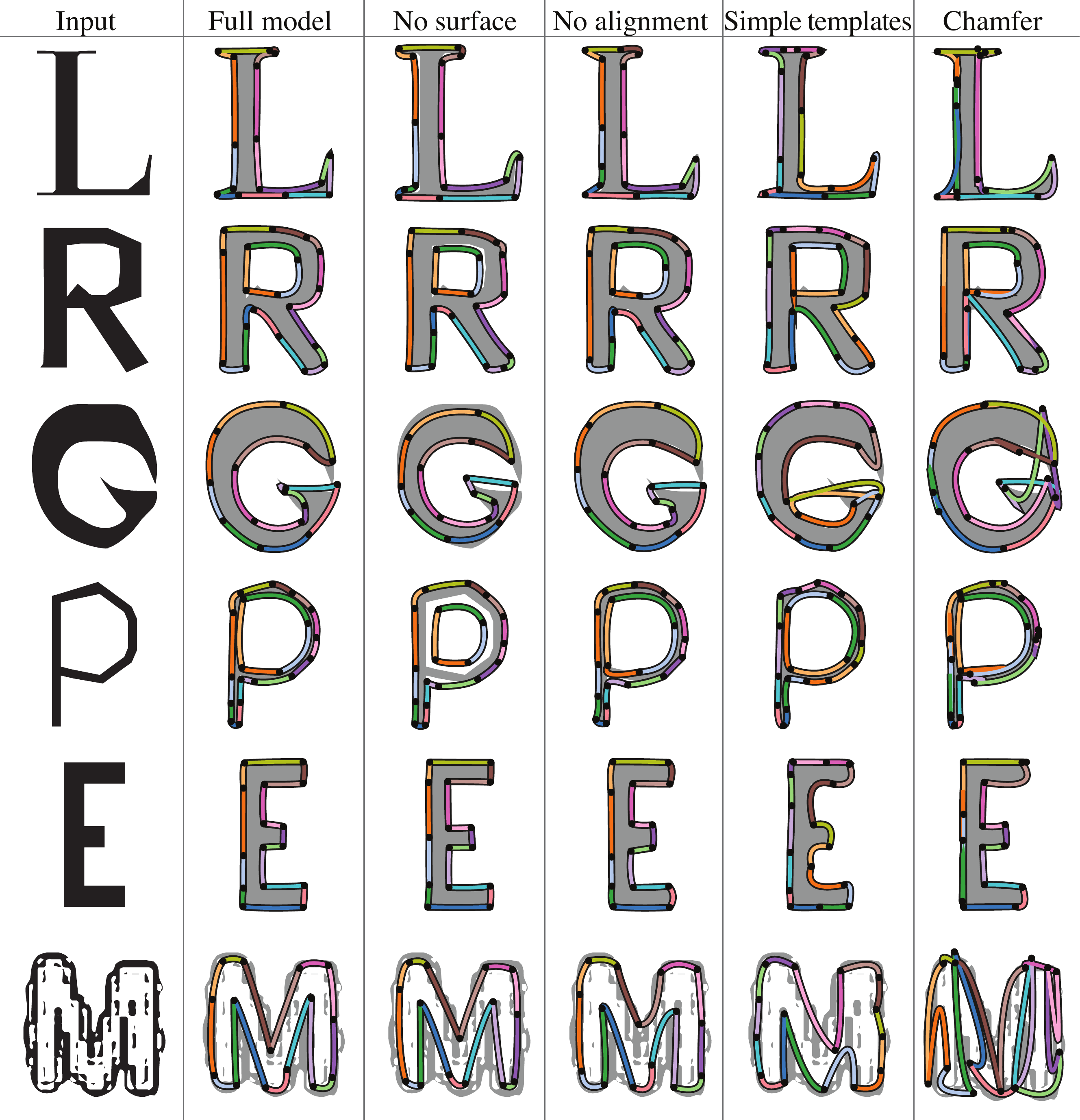}
    \caption{Distance field loss comparisons.}
    \label{fig:sup-ablation}
\end{figure*}
\begin{figure*}[t!]
    \centering
    \includegraphics[width=0.9\linewidth]{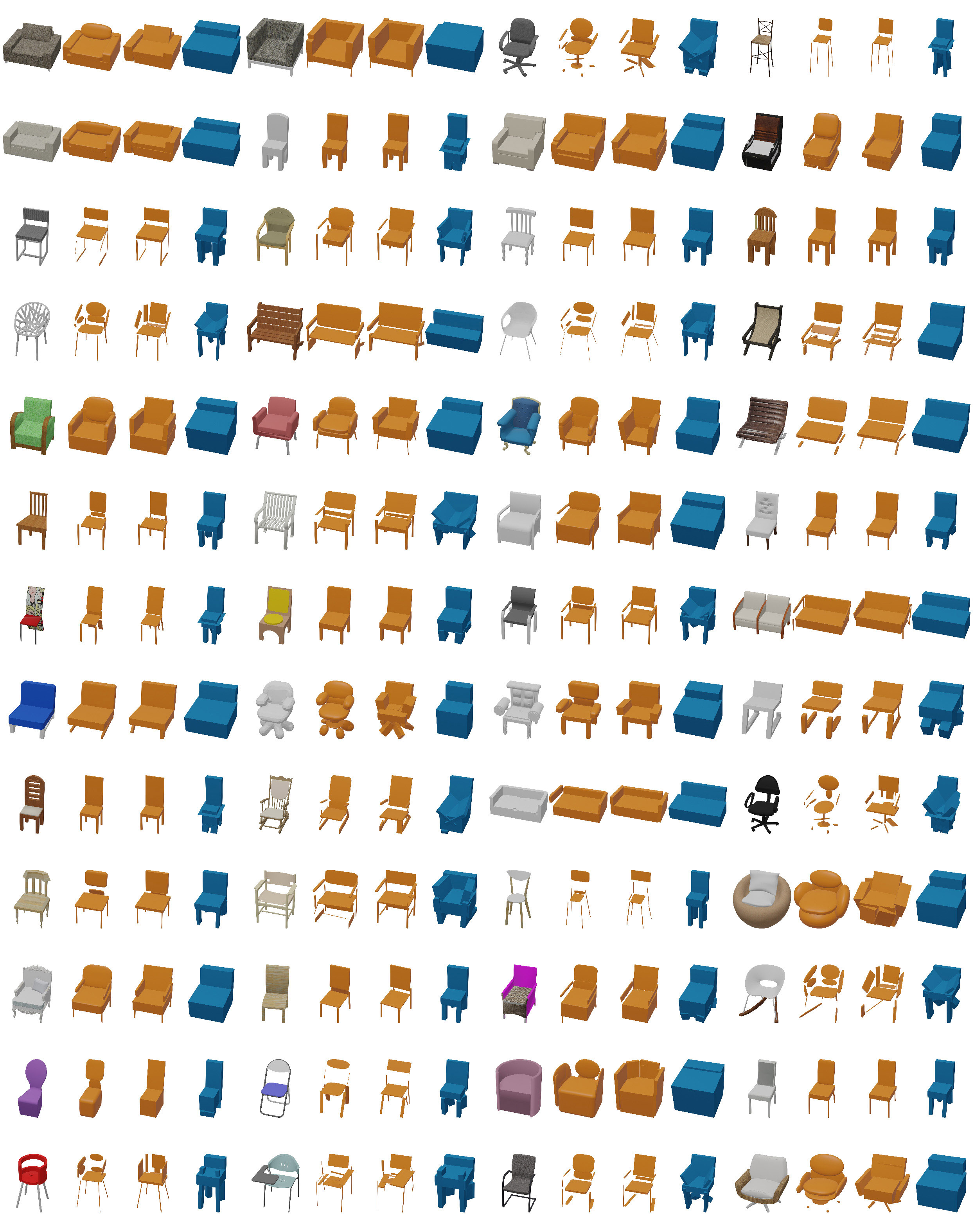}
    \label{fig:sup-chairs}
    \caption{Abstractions of ShapeNet chairs. From left to right, we show the input, our rounded cuboid abstraction, our cuboid abstraction, and the cuboid abstraction of \cite{tulsiani2017learning}}
\end{figure*}
\begin{figure*}[t!]
    \centering
    \includegraphics[width=0.9\linewidth]{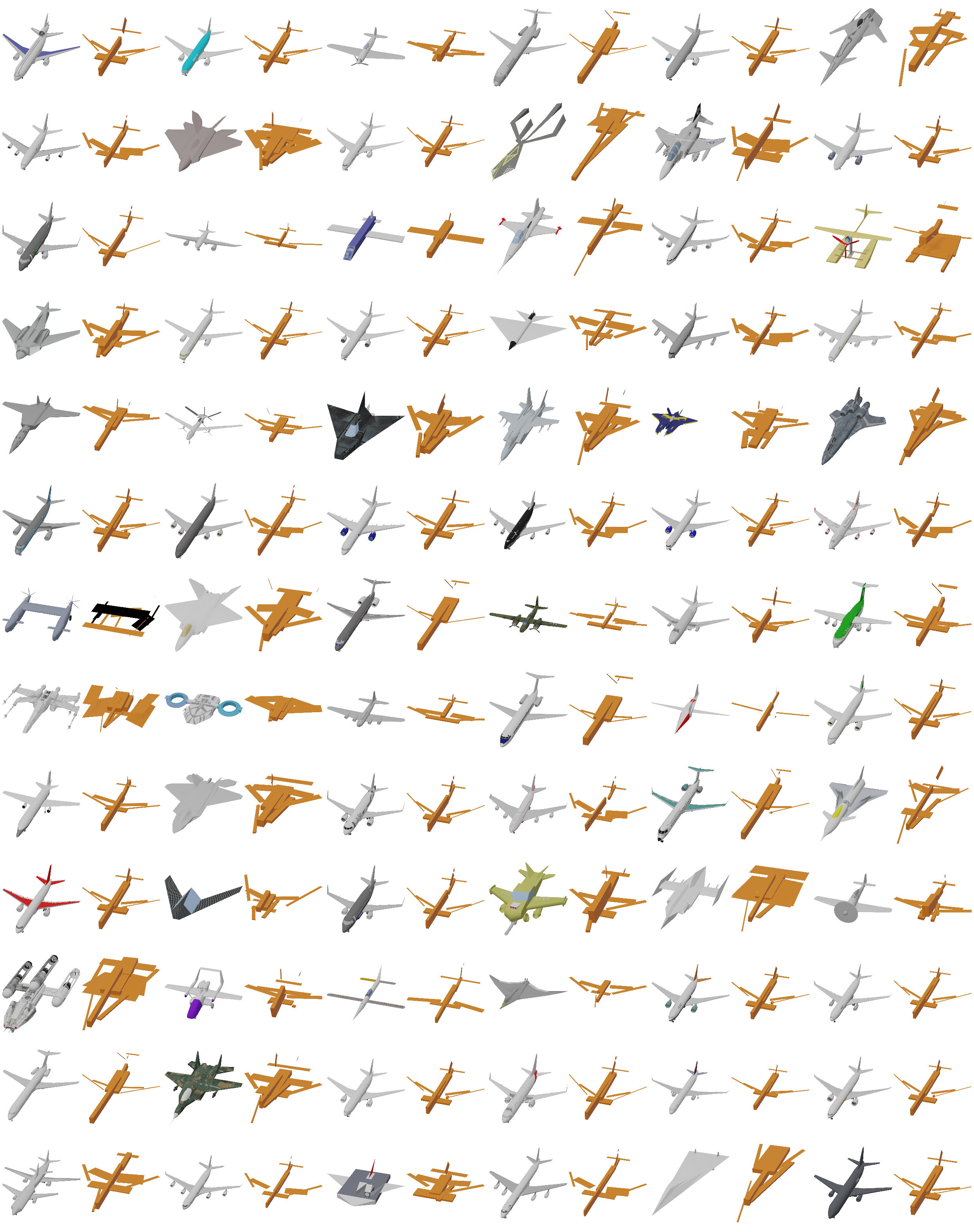}
    \label{fig:sup-airplanes}
    \caption{Cuboid abstractions of ShapeNet airplanes. We show our abstraction (orange) next to each input.}
\end{figure*}